%% file: ms.tex
\documentclass{vldb}
\pdfoutput=1
\usepackage{graphicx}
\usepackage{balance}
\allowdisplaybreaks

\vldbTitle{Revenue Maximization for Query Pricing}
\vldbAuthors{Chawla et al}
\vldbDOI{ https://doi.org/10.14778/3357377.3357378}
\vldbVolume{13}
\vldbNumber{1}
\vldbYear{2019}

\input{preamble}

\input{macros}

\hypersetup{colorlinks,linkcolor={red},citecolor={red}}  
\begin{document}
	\title{Revenue Maximization for Query Pricing}
	
	\numberofauthors{1}
	
	\author{
		\alignauthor Shuchi Chawla, Shaleen Deep, Paraschos Koutris, Yifeng Teng  \\
		\affaddr{University of Wisconsin-Madison} \\
		\affaddr{Madison, WI, USA} \\
		\email{ $\bigl\{$shuchi, shaleen, paris, yifengt$\bigr\}$@cs.wisc.edu}
	}
	
	\maketitle
	
	\input{abstract}	
	\input{introduction_new}

	\input{related}
	\input{querypricing}

	\input{itembundlepricing}

	\input{approxalgo}
	\input{experiments}

	\input{lessons}
	\input{conclusion}

	\bibliographystyle{abbrv}
	\bibliography{qpricing,reference}
	\newpage
	\onecolumn
	\appendix
	\input{appendix}

\end{document}

%% file: preamble.tex
\usepackage{enumitem}
\usepackage{caption}
\usepackage{graphicx}   
\usepackage{hyperref}   
\usepackage{amsmath,amssymb}    
\usepackage{thm-restate}
\usepackage{mdwlist}
\usepackage{mathtools}
\usepackage{xspace}
\usepackage{verbatim}   
\usepackage{xcolor}
\usepackage{makecell}
\usepackage[mathscr]{euscript}
\usepackage{tikz}
\usepackage{booktabs}
\usepackage{tikz-qtree}
\usepackage{subcaption}
\usepackage{multirow}
\usetikzlibrary{arrows.meta}
\usetikzlibrary{fit,shapes,trees,shapes.geometric}
\usetikzlibrary{patterns,decorations.pathreplacing,calc}
\usetikzlibrary{matrix, positioning, arrows}
\usepackage[linesnumbered,algoruled, lined, noend]{algorithm2e}

%% file: macros.tex
\newcommand{\introparagraph}[1]{\smallskip \noindent\textbf{#1.}}   
\newcommand{\cut}[1]{}
\newcommand{\highlight}[1]{{\color{black} #1}}
\newcommand{\camerareadyhighlight}[1]{{\color{black} #1}}

\newcommand{\hhlline}[3]{\draw (#1-#2-1.south west) -- (#1-#2-#3.south east);}

\DeclareMathOperator*{\argmin}{\arg\!\min}
\usepackage{arydshln}
\newcolumntype{L}[1]{>{\raggedright\let\newline\\\arraybackslash\hspace{0pt}}m{#1}}

\newenvironment{packed_item}{
\begin{itemize}
   \setlength{\itemsep}{1pt}
   \setlength{\parskip}{0pt}
   \setlength{\parsep}{0pt}
}
{\end{itemize}}

\newcommand{\ie}{{\em i.e.}\xspace}

\newcommand{\bQ}[0]{\mathbf{Q}}       
       
\newcommand{\bR}[0]{\mathbf{R}}

\newcommand{\bA}[0]{\mathbf{A}}     
\newcommand{\bB}[0]{\mathbf{B}}

\newcommand{\mS}[0]{\mathcal{S}} 
\newcommand{\mI}[0]{\mathcal{I}}   
   
\newcommand{\mH}[0]{\mathcal{H}}   
\newcommand{\mE}[0]{\mathcal{E}}   
\newcommand{\mV}[0]{\mathcal{V}}   
\newcommand{\mC}[0]{\mathcal{C}}      
   
\newcommand{\mL}[0]{\mathcal{L}}

\newcommand{\dagr}[2]{{\mC}_{#1}(#2)} 
\newcommand{\setof}[2]{\{{#1}\mid{#2}\}}
\newcommand{\set}[1]{\{#1\}}   
\newcommand{\qb}[1]{\langle {#1} \rangle}   

\newcommand{\db}{\mathcal{D}}%

\newcommand{\ubp}{\textsf{UBP}}
\newcommand{\uip}{\textsf{UIP}}
\newcommand{\lpip}{\textsf{LPIP}}
\newcommand{\cip}{\textsf{CIP}}
\newcommand{\xos}{\textsf{XOS}}
\newcommand{\ssb}{\texttt{\bfseries SSB}}
\newcommand{\tpch}{\texttt{\bfseries TPC-H}}

\definecolor{Gray}{gray}{0.9}
\definecolor{cardinal}{rgb}{0.77, 0.12, 0.23}
\definecolor{bondiblue}{rgb}{0.0, 0.58, 0.71}
\definecolor{babyblue}{rgb}{0.54, 0.81, 0.94}
\definecolor{lightgreen}{rgb}{0.67, 0.94, 0.82}
\definecolor{maize}{rgb}{0.98, 0.93, 0.37}

\hypersetup{colorlinks,citecolor={cardinal}}

\newtheorem{example}{Example}

\newtheorem{theorem}{Theorem}


%% file: abstract.tex
\begin{abstract}
Buying and selling of data online has increased substantially over the last few years. Several frameworks have already been proposed that study query pricing in theory and practice. The key guiding principle in these works is the notion of {\em arbitrage-freeness} where the broker can set different prices for different queries made to the dataset, but must ensure that the pricing function does not provide the buyers with opportunities for arbitrage. However, little is known about revenue maximization aspect of query pricing. In this paper, we study the problem faced by a broker selling access to data with the goal of maximizing her revenue. We show that this problem can be formulated as a revenue maximization problem with single-minded buyers and unlimited supply, for which several approximation algorithms are known. We perform an extensive empirical evaluation of the performance of several pricing algorithms for the query pricing problem on real-world instances. In addition to previously known approximation algorithms, we propose several new heuristics and analyze them both theoretically and experimentally. Our experiments show that algorithms with the best theoretical bounds are not necessarily the best empirically. We identify algorithms and heuristics that are both fast and also provide consistently good performance when valuations are drawn from a wide variety of distributions.
\end{abstract}

%% file: introduction_new.tex
\section{Introduction}
\label{sec:intro}

The last decade or so has seen an explosion of data being collected from a variety of sources and across a broad range of areas. Many companies, including Bloomberg~\cite{bloomberg}, Twitter~\cite{twitterapi}, Lattice Data~\cite{lattice}, DataFinder~\cite{datafinder}, and Banjo~\cite{banjo} collect such data, which they then sell as structured (relational) datasets. 
These datasets are also often sold through online {\em data markets}, which are web platforms for buying and selling data: examples include BDEX~\cite{bdex}, Salesforce~\cite{salesforce} and QLik DataMarket~\cite{qlik}. Even though data sellers and data markets offer an abundance of data products, the pricing schemes currently used are very simplistic. In most cases, a data buyer has only one option, to buy the whole dataset at a fixed price. Alternatively, the dataset is split into multiple disjoint chunks, and each chunk is sold at a separate price.

However, buyers are often interested in extracting specific information from a dataset and not in acquiring the whole dataset. Accessing this information can be concisely captured through a {\em query}. Selling the whole dataset at a fixed price forces the buyer to either pay more for the query than it is valued, or to not buy at all. This means that valuable data is often not accessible to entities with limited budgets, and also that data-selling companies and marketplaces behave suboptimally with respect to maximizing their revenue. \camerareadyhighlight{Indeed, popular cloud database providers such as Google~\cite{google} and Amazon~\cite{amazon} also follow a coarse grained pricing model where the user is charged based on the number of bytes scanned rather than information content of the requested query.}

\introparagraph{Query-Based Pricing}
To address this problem, a recent line of research~\cite{KUBHS12,KUBHS13,deep2017qirana} introduced the framework of query-based pricing. A {\em query-based pricing scheme} tailors the purchase of the data to the user's needs, by assigning a different price to each issued query. Given a dataset $\db$ and a query $Q$ over the dataset, the user must pay a price $p(Q,\db)$ to obtain the answer $Q(\db)$. This price reflects only the value of the information learned by obtaining the query answer, and not the computational cost of executing the query. The work on query-based pricing has mainly focused on how to define a well-behaved pricing function, and how to develop system support for efficiently implementing a data marketplace.
In particular, a key property that a pricing function must obey is that of {\em arbitrage-freeness}: it should not be possible for the buyer to acquire a query for a cheaper price through the combination of other query results. The arbitrage-freeness constraint makes the design of pricing functions a challenging task, since deciding whether a query is more informative than another query (or set of queries) is generally computationally hard, and for practical applications it is critical that the price computation can be performed efficiently.

To overcome this barrier, \cite{deep2017qirana} proposes a setup where we start with a set $\mS$ consisting of multiple  "candidate" databases instances; this set is called the {\em support}.
Each query $Q$ can then be thought of as a function that classifies instances from $\mS$: ones that return the same  answer as $Q(\db)$, and ones that do not.
Whether a query is more informative than another then amounts to whether it classifies more inconsistent instances than the other. The benefit of this approach is that in order to find the price of a query, it suffices to only examine the instances in the set $\mS$, which is a computationally feasible problem.

\introparagraph{The Revenue Maximization Problem}
Although prior work provides a framework to reason about the formal properties of pricing functions, it does not address the following fundamental question:

\vspace{0.5em}
\noindent{ \em How do we assign prices to the queries in order to maximize the seller's revenue while ensuring arbitrage freeness?}
\vspace{0.5em}

This is the main problem we study in this paper. 
Our key observation is that query pricing can be cast as a problem of pricing subsets ({\em bundles}) over a ground set of {\em items}, where each item corresponds to a database instance in $\mS$. The arbitrage-freeness constraint corresponds to the pricing function (which is a set function over $\mS$) being {\em monotone} and {\em subadditive}. 
Since there is no limit to how many times a seller can sell a query (a digital good), we can model the seller as having {\em unlimited supply} for each query answer. We further consider {\em single minded} buyers, which means that each buyer wants to buy the answer to a single set of queries. 

Finding the monotone and subadditive pricing function that maximizes revenue in this setting is a computationally hard problem. Furthermore, a subadditive function, even if we manage to find one, can take exponential space (w.r.t. $\mS$) to store. Therefore, {\em for practical applications we must seek a simple and concise pricing function that approximates the optimal subadditive pricing in terms of revenue}. 
In this paper, we explore such succinct families of pricing functions that are appropriate for use in a data market, and
answer the following questions:
\begin{itemize}
\item What is the theoretical gap between optimal revenue and the revenue obtained through succinct families of pricing functions?
\item Which revenue maximization algorithms are best suited for query pricing and what guarantees do they offer?
\item How well do the theoretical revenue and performance bounds translate to real-world query workloads? 
\end{itemize}

\introparagraph{Our Contributions}
We now discuss our contributions in detail.

\smallskip
\noindent \textit{Succinct Pricing Functions.\:}
We study three types of succinct pricing functions (Section~\ref{sec:itembundlepricing}). The first, {\em uniform bundle pricing}, assigns the same price to every bundle (query) and is the default pricing scheme in many data markets. 
The second, additive or {\em item pricing}, assigns a price to each item (instance in the support) and charges a price for each bundle equal to the sum of prices for the items in the bundle.  Item pricing has been studied extensively from a theoretical perspective and a number of approximation algorithms are known (see, e.g.,\cite{guruswami2005profit,balcan2006approximation,briest2006single}). 
Third, we consider a much more general class of pricings, namely {\em \xos\ or fractionally subadditive pricings}. These pricing functions are more expressive than item or uniform bundle pricings, while at the same time having a small representation size. 
Our key finding is that \xos\ pricing functions can achieve a logarithmic factor larger revenue than the better of item and uniform bundle pricing.

\smallskip
\noindent \textit{Revenue Maximization Algorithms.\:}
We theoretically study several algorithms for finding the revenue maximizing pricing function (Section~\ref{section-approxalgo}). In quantifying performance, several parameters of the instance are relevant: the number of items $n$ (which is the size of the support), the number of bundles $m$ (which is the number of the queries issued in the market), the size of the largest bundle $k$, and the maximum number of bundles any item belongs to $B$. In the context of query pricing, it is usually the case that $B\le m\ll k\le n$, so algorithms with approximation factors and running time depending on $B$ or $m$ are generally better than those depending on $k$ or $n$.

In particular, although we can always find the optimal uniform bundle pricing efficiently, it is computationally hard to find the optimal item pricing. Hence, we consider several algorithms for the latter task that come with worst case approximation factors that are logarithmic in one or more of the natural parameters of the instance. Apart from known algorithms, we also develop new algorithmic techniques that improve performance.
Finally, for the family of \xos\ pricing functions, we propose an algorithm that simply combines multiple additive item pricing functions. 


\smallskip
\noindent \textit{Experimental Evaluation.\:}
 Finally, we perform an empirical analysis of the different pricing functions to understand how well do the algorithms hold up in practice, which of these algorithms should a practitioner use, and what features of the problem instance dictate this choice (Section~\ref{sec:experiments}).
We compare the pricing algorithms using both synthetic and real world query workloads.  

Our study shows that the worst-case analysis of pricing algorithms does not capture how well the algorithms behave in real-world instances in terms of approximating the optimal revenue. In particular, we observe that the structure of the bundles induced by different query workloads, as well as the distribution of buyer valuations, heavily influences the quality of approximation. For example, the algorithm that obtains the best known worst case approximation ratio does not achieve the best performance of the algorithms we tested in any of our setups. 
Our experiments also show that it is possible to efficiently extract most of the available revenue using succinct pricing functions, and in particular item pricings. Hence, succinct pricing functions seem a good practical choice for a data marketplace.

\smallskip
\noindent \textit{Lessons and Open Problems.\:}
Finally, we discuss our takeaway lessons, as well as several exciting open questions (Section~\ref{sec:lessons}).

%% file: related.tex
\section{Related Work}
\label{sec:related}

\introparagraph{Query-Based Pricing}
There exist various simple mechanisms for data pricing (see~\cite{muschalle2012pricing} for a survey on the subject), including a flat fee tariff, usage-based and output-based prices. 
These pricing schemes do not provide any guarantees against arbitrage.
The vision for arbitrage-free query-based pricing was first introduced by Balazinska et al.~\cite{balazinska:11b}, and was further developed in a series of papers~\cite{KUBHS12,KUBHS13,KUBHS12b}. The proposed framework requires that the seller sets fine-grained price points, which are prices assigned to a specific type of queries
over the dataset; these price points are used as a guide to price the incoming queries. 
Even though the pricing problem in this setting is in general NP-hard, the QueryMarket prototype~\cite{KUBHS13} showed that it is feasible to compute the prices for small datasets, albeit not in real-time or maximizing the revenue.  Further work on data pricing proposed new criteria for interactive pricing in data markets~\cite{LM12}, and described new necessary arbitrage conditions along with several negative results related to the trade-off between flexible pricing and arbitrage avoidance~\cite{LK14}. Upadhyaya et al.~\cite{Upadhyaya2016} investigated history-aware pricing using refunds. More recently,~\cite{deep2016design} characterized the possible space of pricing functions with respect to different arbitrage conditions. The theoretical framework was then implemented as part of the \textsc{Qirana} system~\cite{deep2017qirana, deep2017qiranademo}, which can support pricing in real-time. The framework we use in this paper for query-based pricing is the same one from~\cite{deep2017qirana}.

\introparagraph{Revenue Maximization}
Revenue-maximizing mechanisms have been well understood in single-item auctions, where the posted pricing
mechanism is optimal \cite{myerson1981optimal}. 
However, in general multi-parameter settings, revenue-maximizing mechanisms are considered
hard to characterize. In the past few decades, many researchers started to 
focus on simple and approximately optimal solutions, especially posted-pricing mechanisms.
Recent line of work shows that in Bayesian setting with limited supply, posted pricing achieves 
constant approximation when there is single buyer 
\cite{babaioff2014simple, chawla2007algorithmic, 
chawla2010multi, chawla2015power, rubinstein2015simple},
and logarithmic approximation (with respect to the number of items) when there are multiple buyers 
\cite{cai2016duality, chawla2016mechanism, caizhao2017duality}.

In this paper, we focus on the case where all valuations of buyers are revealed to the seller. The study of this setting was initiated by \cite{guruswami2005profit}, which shows that item pricing 
gives $O(\log n)$-approximation for unit-demand buyers
in limited-supply setting, and $O(\log n+\log m)$-approximation for single-minded buyers in unlimited supply
setting. The competitive ratio for unlimited supply setting was improved to $O(\log k + \log B)$ 
by \cite{briest2006single} then to $O(\log B)$ by \cite{cheung2008approximation} where 
$k$ denotes the size of largest bundle, and $B$ denotes the maximum number of bundles containing a specific item.
Another line of work studies how to find best possible item pricing in above setting where $k$ is bounded. This problem 
is also known as the \textit{k-hypergraph pricing} problem.
\cite{briest2006single} gave the first polynomial-time algorithm finding an approximately optimal item pricing
with competitive ratio $k^2$. The approximation ratio is improved to $k$ by \cite{balcan2006approximation},
which is proven to be near-optimal:  under the
Exponential Time Hypothesis there is no polynomial-time
algorithm that achieves competitive ratio $k^{1 - \epsilon}$~\cite{chalermsook2013independent}.

\introparagraph{Pricing information}
Another line of research in economics considers the revenue maximization problem for a seller offering to sell information. See, for example, \cite{BKP-12, BB-15, BBS-17} and references therein. However, that literature differs from our work in several fundamental aspects. First, in those works, both the seller and the buyer are unaware of the true state of the information (\ie, the dataset), and this state is stochastic. Second, the seller is allowed to sell queries whose results are randomized. Third, the buyer's type (which information he is interested in and his value) are unknown to the seller. In contrast, in our setting while the pricing is required to be arbitrage-free, the types of the buyers are known to the seller in advance. As such the two models lead to very different types of pricing mechanisms and algorithms.

%% file: querypricing.tex
\section{The Query-Based Pricing Framework}
\label{sec:framework}

In this section, we present the framework of query-based pricing proposed by \cite{deep2016design}, and then formally describe the pricing problems we tackle.

\subsection{Query-Based Pricing Basics}

The {\em data seller} wants to sell an instance $\db$ through a data market, which functions as the broker. \textcolor{black}{The instance has a fixed relational schema $\bR = (R_1, \dots, R_k)$.} We denote by $\mI$ the set of possible database instances. The set $\mI$ encodes information about the data that is provided by the data seller, and is public information known to any buyer (together with the schema). We allow the set $\mI$ to be infinite, but countable. \textcolor{black}{For example, suppose that the schema consists of a single binary relation $R(A,B)$, and the domain of both attributes is $[\ell] = \set{1, \dots, \ell}$. Then, $\mI = 2^{[\ell] \times [\ell]}$, \ie the set of all directed graphs on the vertex set $[\ell]$.}

{\em Data buyers} can purchase information by issuing queries on $\db$ in the form of a {\em query vector}  $\bQ=\qb{Q_1, \hdots, Q_p}$. For our purposes, a query $Q$ is a deterministic function that takes as input a database instance $\db$ and returns an output $Q(\db)$. We denote the output of the query vector by $\bQ(\db) = \qb{Q_1(\db), \hdots Q_p(\db)}$. 

\begin{example}
	\textit{Consider a database that consists of a single \textsc{User} relation as shown in Figure~\ref{fig:example}. Suppose that the data seller has fixed the price of the entire relation to $\$100$. Consider a data buyer, Alice, who is a data analyst and wants to study user demographics. Since Alice has a limited budget, she cannot afford to purchase the entire database. Thus, Alice will extract information from the table by issuing relational queries over time. We will use this as a running example throughout the section.}
\end{example}

\begin{figure}[t]
	\centering
			\begin{tabular} {l|l|c|c}
				\multicolumn{4}{ l }{\textsc{User}} \\ 
				\hline
				\textsf{\underline{uid}} & \textsf{name} & \textsf{gender} & \textsf{age}  \\ \hline
				1 & Abe &  m & 18 \\ 
				2 & Alice  & f & 20 \\
				3 & Bob & m & 25 \\
				4 & Cathy & f & 22 \\
				\hline
			\end{tabular}
			\caption{A relation with $4$ attributes. \textsf{\underline{uid}} is the primary key.}
			\label{fig:example}
\end{figure}

A {\em pricing function} $p(\bQ, \db)$ takes as input a query vector $\bQ$ and a database instance $\db \in \mI$ and assigns to it a price,\footnote{Allowing prices to be general functions of query vectors, rather than just additive over queries allows for more expressivity and therefore more revenue for the seller.} which is a number in $ \mathbb{R}_+$. 
Assigning prices to query vectors without any restrictions can lead to arbitrage opportunities in the following two ways:

\introparagraph{Information Arbitrage} The first condition captures the intuition that if a query vector $\bQ_1$ reveals a subset of information of what a query vector $\bQ_2$ reveals, then the price of $\bQ_1$ must be no more than the price of $\bQ_2$. If this condition is not satisfied, it creates an arbitrage opportunity, since a data buyer can purchase $\bQ_2$ instead, and use it to obtain the answer of $\bQ_1$ for a cheaper price. 

Formally, we say that $\bQ_2$ determines $\bQ_1$ under database $\db$ if for every database $\db' \in \mI$ such that $\bQ_2(\db) = \bQ_2(\db')$, we also have  $\bQ_1(\db') = \bQ_1(\db)$.
We say that the pricing function $p$ has no {\em information arbitrage} if for every database $\db \in \mI$ such that $\bQ_2$ determines $\bQ_1$ under $\db$, we have $p(\bQ_2, \db) \geq  p(\bQ_1, \db)$.

\begin{example} \label{ex:2}
	\textit{In our running example, suppose that Alice wants to count the number of female users in the relation. She can issue the query $Q_1 = \texttt{SELECT count(*) FROM User}$ $\texttt{WHERE gender = 'f'}$. However, a different way she can learn the same information is by issuing the query $Q_2 = \texttt{SELECT gender,}$ $\texttt{ count(*) FROM User }$ $\texttt{GROUP BY gender}$. $Q_2$ will return the number of users for each gender as its output. Suppose that the buyer charges $p(Q_1) = \$10$ and $p(Q_2) = \$5$, then there exists an information arbitrage opportunity. Since Alice can learn the required information from $Q_2$ at a {\emph{cheaper}} cost, she has no incentive to purchase $Q_1$. Thus, if the seller wants to prevent this arbitrage, he needs to ensure that $p(Q_1) \leq p(Q_2)$. Thus, the seller now sets the price of $Q_2$ to $\$10$, i.e, $p(Q_2) = \$10$.}
\end{example}

\introparagraph{Combination Arbitrage} The second condition regards the scenario where a data buyer wants to obtain the answer for the query vector $\bQ = \bQ_1\Vert \bQ_2$, where $\Vert$ denotes vector concatenation. Instead of asking $\bQ$ as one, the buyer can create two separate accounts, and use one to ask for $\bQ_1$ and the other to ask for $\bQ_2$. To avoid such an arbitrage situation, we must make sure that the price of $\bQ$ is at most the sum of the prices for $\bQ_1$ and  $\bQ_2$. 
Formally, we say that the price function $p$ has no {\em combination arbitrage} if for every database $\db \in \mI$, we have $p(\bQ_1\Vert \bQ_2, \db) \leq p(\bQ_1, \db) + p(\bQ_2, \db)$.

\begin{table}[t]
	\small
	\centering
	\caption{Symbol definitions.}
	\hspace*{-1.5mm}
	\scalebox{1}{
	\begin{tabular}{p{1.5cm}lp{0.2cm}}
		\toprule
		\textbf{Symbol}                    &{\textbf{Description}}                                                                                                  \\ 
		\midrule 
		$\bQ$    & Query vector containing buyer queries  \\ 
		$\db$      & Input database provided by the seller \\ 
		$\mI$       & \begin{minipage}[t]{0.73\columnwidth}Set of all possible database instances consistent with $\db$\end{minipage} \\
		$\mS$      & Support set chosen by the framework \\ 
		$p(\bQ,\db)$      & Pricing function\\ 
		$\dagr{\mS}{\bQ,\db}$      & Conflict set of query $\bQ$ (subset of $\mS$) \\ 
		$\mV$      &  Vertex set of hypergraph \\ 
		$\mE$      &  Hyperedges in a hypergraph \\
		$\mH = (\mV, \mE)$     &  \begin{minipage}[t]{0.73\columnwidth}%
			Hypergraph generated by transforming buyer queries into hyperedges (bundles) containing db's in conflict set ($\mV = \mS$)\end{minipage} \\
		item $j$ &  some database $j \in \mV$ in vertex set of $\mH$ \\
		$v_{\bQ}$      & buyer valuation of query vector $\bQ$ \\ 
		bundle $A$ & Conflict set of some query in $\mH$. \\ 
		$B$ & maximum degree over all items in $\mH$ \\
		\bottomrule
	\end{tabular}
	}
\label{tab:symbols}
\end{table}

\begin{example}
	\textit{Alice now wants to find the average age of female users in the relation. She can issue the query $Q_3 = \texttt{SELECT AVG(age)}$  $\texttt{FROM User WHERE gender = 'f'}$. Suppose the seller decides to price $p(Q_3) = \$20$. However, Alice could have also chosen to ask $Q_4 = \texttt{SELECT SUM(age) FROM User}$ $\texttt{ WHERE gender = 'f'}$. Now, she can obtain her desired result by combining the answers of $Q_4$ and $Q_2$ \footnote{We assume that it is public information that gender in this relation takes only two values: \textsf{m} and \textsf{f} }. If the seller prices $p(Q_4) = \$5$, then there exists a combination arbitrage opportunity, since $p(Q_3) > p(Q_4) + p(Q_2)$ which gives Alice an incentive to split her query. To avoid this, the seller needs to ensure that $p(Q_3) \leq p(Q_4) + p(Q_2)$.}
\end{example}

We say that the pricing function $p$ is {\em arbitrage-free} if it has no information arbitrage and no combination arbitrage. 

\subsection{From Pricing Queries to Pricing Bundles}

In general, computing whether $\bQ_2$ determines $\bQ_1$ under some $\db$ is an intractable problem. To overcome this obstacle, we take a different view of a query vector. Let 
$\mS \subseteq \mI$ be any subset of $\mI$, called the {\em support}, and define the {\em conflict set} of $\bQ$ with respect to $\mS$ as:
\begin{align*}
\dagr{\mS}{\bQ,\db}  &= \setof{\db' \in \mS}{\bQ(\db) \neq \bQ(\db')}. 
\end{align*}

Intuitively, the conflict set contains all the instances from $\mS$ for which the buyer knows that cannot be the underlying instance $\db$ once she learns the answer $\bQ(\db)$. This construction maps each query vector to a {\em  bundle} $\dagr{\mS}{\bQ,\db} $ over the set $\mS$. We should remark here that the task of computing the bundle $\dagr{\mS}{\bQ,\db} $ is computationally feasible if we choose $\mS$ to be small enough, since we can simply iterate through all the items $\db' \in \mS$, and for each item check the condition 
$\bQ(\db) \neq \bQ(\db')$.

\begin{example} \label{ex:conflictset}
	
	Consider the support set $\mS = \{\db_1, \db_2, \db_3\}$ as shown below. The colored font highlights the values that are changed with respect to $\db$.
	
	\begin{tikzpicture}
	\matrix   [matrix of nodes,
	row sep=-\pgflinewidth,
	column sep=-\pgflinewidth,
	nodes={rectangle,minimum width=2em, outer sep=0pt},
	scale=0.7, every node/.style={scale=0.7}] at (-1.5,-4) (db1)
	{
		\textsf{\underline{uid}} & \textsf{name} & \textsf{gender} & \textsf{age} \\
		1   & Abe & m & 18\\
		2 & Alice  & f & {\color{blue} \bf 30} \\
		3 & Bob & m & 25 \\
		4 & Cathy & f & 22 \\
	};
	\hhlline{db1}{1}{4};
	
		\matrix   [matrix of nodes,
	row sep=-\pgflinewidth,
	column sep=-\pgflinewidth,
	nodes={rectangle,minimum width=2em, outer sep=0pt},
	scale=0.7, every node/.style={scale=0.7}] at (1.3,-4) (db2)
	{
		\textsf{\underline{uid}} & \textsf{name} & \textsf{gender} & \textsf{age} \\
		1   & Abe & {\color{blue} \bf  f} & 18\\
		2 & Alice  & f & 20 \\
		3 & Bob & m & 25 \\
		4 & Cathy & f & 22 \\
	};
	\hhlline{db2}{1}{4};
	
	\matrix   [matrix of nodes,
	row sep=-\pgflinewidth,
	column sep=-\pgflinewidth,
	nodes={rectangle,minimum width=2em, outer sep=0pt},
	scale=0.7, every node/.style={scale=0.7}] at (4.1,-4) (db3)
	{
		\textsf{\underline{uid}} & \textsf{name} & \textsf{gender} & \textsf{age} \\
		1   & Abe & m & 18\\
		2 & Alice  & f & 20 \\
		3 & {\color{blue} \bf  Ben} & m & 25 \\
		4 & Cathy & f & 22 \\
	};
	\hhlline{db3}{1}{4};
	
	\node[above=0pt of db1]
	(cell1) {$\db_1$};
	
	\node[above=0pt of db2]
	(cell2) {$\db_2$};
	
	\node[above=0pt of db3]
	(cell3) {$\db_3$};
	\end{tikzpicture}
	
For query $Q_1$ from Example~\ref{ex:2}, $Q_1(\db) = Q_1(\db_1)$ $= Q(\db_3)$ but $Q_1(\db) \neq Q_1(\db_2)$. Thus, $\dagr{\mS}{Q_1,\db} = \{\db_2\}$. Similarly, $\dagr{\mS}{Q_3,\db} = \{\db_1, \db_2\}$.
\end{example}

We can now compute a price for $\bQ$ by applying a set function $f: 2^{\mS} \rightarrow  \mathbb{R}_+$ to $\dagr{\mS}{\bQ, \db}$. 
A set function $f$ is {\em monotone} if for sets $A \subseteq B$ we always have $f(A) \leq f(B)$, and {\em subadditive} if for every set $A,B$ we have $f(A) + f(B) \geq f(A \cup B)$.
By choosing $f$ to be monotone and subadditive, we can guarantee that the pricing function is arbitrage-free.

\begin{theorem}[\protect\cite{deep2016design}] \label{cor:arbitrage}
Let $\mS \subseteq \mI$, and $f$ be a set function $f: 2^{\mS} \rightarrow  \mathbb{R}_+$. Then, the pricing function $p(\bQ, \db) = f(\dagr{\mS}{\bQ, \db})$ is arbitrage-free
if and only if the function $f$ is monotone and subadditive. 
\end{theorem}

We emphasize that the arbitrage-freeness guarantee holds for any support $\mS$. The choice of $\mS$ impacts the granularity of prices that can be assigned to queries which in turn affects the revenue. Observe that in the extreme case when $\mS = \emptyset$, $\dagr{\mS}{\bQ,\db} = \emptyset$ for any $\bQ$ implying that all queries have exactly the same price. On the other hand, a large support set can make the computation of the conflict set prohibitively expensive. Section~\ref{sec:varying:support} explores the tradeoff between the revenue obtained and $\mS$ in more detail.

\subsection{Revenue Maximization}

We consider the unlimited supply setting, where the seller can sell any number of units of each query. Additionally, we assume that the buyers are single-minded: each buyer is interested in buying only a single query vector $\bQ$; the buyer will purchase $\bQ$ only if $p(\bQ, \db) \leq v_\bQ$, where $v_\bQ$ is the valuation  that the buyer has for $\bQ$. \camerareadyhighlight{Note that the single-minded buyer assumption is not restrictive; a buyer who wishes to purchase multiple queries (say $Q_1, \dots, Q_\lambda$) can be modeled as $\lambda$ separate buyers where each buyer $i \in [\lambda]$ wants to purchase query $\bQ_i = \qb{Q_i}$. Similarly, if the buyer wants to buy all $\lambda$ queries together, then we can already express it as a bundle $\bQ = \qb{Q_1, \dots, Q_\lambda}$.}

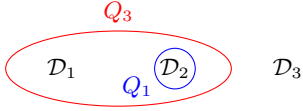
\begin{figure} \centering
	\begin{tikzpicture}
		\node (item1)  at (1,0) {$\db_1$};
		\node (item2)  at (2.5,0) {$\db_2$};
		\node (item3)  at (4,0) {$\db_3$};
		\draw[color=red] (1.75,0) ellipse (1.5cm and 0.5cm);
		\node at (1.75,0.75) {\color{red}$Q_3$};
		
		\draw[color=blue] (2.5,0) ellipse (0.27cm and 0.27cm);
		\node at (2, -0.25) {\color{blue}$Q_1$};
	\end{tikzpicture}
	\caption{Hypergraph instance for Example~\ref{ex:conflictset}. $Q_1$ and $Q_3$ are represented as hyperedges containing the databases in their conflict sets.}
	\label{fig:hypergraph}
\end{figure}

The problem setup is as follows. We are given as input a set of $m$ buyers, where each buyer $i$ is interested of purchasing a query vector $\bQ_i$ with valuation $v_i$. These valuations can be found by performing market research in order to understand the demand and price buyers are willing to pay for queries of interest. Defining and using these demand curves is a standard practice in the study of economics for digital goods~\cite{gans2011principles}. We pick a support set $\mS \subseteq \mI$ of size $n = |\mS|$. By using the transformation of query vectors to bundles over $\mS$, we can construct a hypergraph $\mH = (\mV,\mE)$, with vertex set $\mV = \mS$, and hyperedges $\mE = \setof{e_i}{i =1, \dots, m}$, where $e_i =\dagr{\mS}{\bQ_i, \db}$. 
Figure~\ref{fig:hypergraph} shows an example of hypergraph instance constructed from the queries and their conflict sets.

A pricing function $p$ is a set function that maps subsets of $S$ to prices in $\mathbb{R}_+$\footnote{Here we have overloaded $p$ to also be a pricing function with input a bundle of items.}. The task at hand is to find a monotone and subadditive pricing $p$ that maximizes the seller's revenue. The revenue of a pricing function $p$ is given by:
\[R(p) = \sum_{i: v_i\ge p(e_i)} p(e_i)\]
The optimal revenue is:
\[\textsc{OPT} = \max_{\text{monotone; subadditive } p} R(p)\]
Many of the approximation results in the literature use a simpler (and weaker) upper bound on $\textsc{OPT}$, namely the sum of all bundle values $\sum_i v_i$, as a basis of comparison for algorithms' performance.  Throughout the paper, we will use the term hypergraph to refer to the instance created by the transformation as described above and the term {\em item} to refer to some instance in the vertex set $\mV$ of the hypergraph.

\subsection{Simple Pricing Functions}

For practical applications (e.g., \textsc{Qirana}~\cite{deep2017qirana}), we must only consider functions that can be both concisely represented and also efficiently computable. For example, it is not desirable to come up with a function $p$ where we need to explicitly store all the $2^n$ values for all input bundles from $\mS$.
For this reason, we focus on a few important subclasses of monotone and subadditive set functions:
\begin{packed_item}
\item The {\em uniform bundle price}  $p^b(\cdot)$ assigns the same price to every hyperedge, \ie $p^b(e) = P$ for some number $P \geq 0$.
\item The {\em additive price} $p^a(\cdot)$ assigns a weight $w_j \geq 0$ to every item $j \in \mS$, and then defines
$p^a(e) = \sum_{j \in e} w_j$. Such a pricing function is also commonly known as an {\em item pricing}.
\item The {\em \xos\ price} $p^x(\cdot)$ defines $k$ weights $w_j^1, w_j^2, \dots, w_j^k$ for each item $j \in \mS$, and sets the price to $p^x(e) = \max_{i=1}^k \sum_{j \in e} w_j^i$.
\end{packed_item}

Given the above three subclasses of pricing functions, we consider the following two questions, both from a theoretical and practical point of view.
First, how much do we lose in terms of revenue by replacing the optimal monotone and subadditive pricing function with a uniform, additive or \xos\ pricing function? In other words, we seek to understand what is the revenue we lose for the sake of computational efficiency.
Second, we want to develop algorithms that can optimize the prices for each subclass and achieve a good approximation ratio with respect to the optimal revenue.

%% file: itembundlepricing.tex
\section{Upper and Lower Bounds}
\label{sec:itembundlepricing}

\begin{figure}[t]
	\centering
	\scalebox{.99}{
		\begin{tikzpicture}
		\tikzset{recstyle/.style={rectangle, draw, minimum width=7mm, minimum height=10mm,rounded corners=.2cm}}
		
		\node[recstyle] (rectbundle) at (0,0.5) {\shortstack{subadditive bundle \\ pricing}};
		\node[recstyle] (rectxos)  at (0,-1.5) {XOS pricing};
		\node[recstyle] (rectmax) at (0,-3.5) {$\max \Big\{  \shortstack{\small item pricing \\ \small uniform bundle pricing} \Big\}$};
		\node[recstyle] (rectitem) at (-2,-5.5) {item pricing};
		\node[recstyle] (rectubundle)  at (2,-5.5) {\shortstack{uniform bundle \\ pricing}};

		\draw [->] (rectubundle.east) to [out=60, in = 0] (rectbundle);
		\node at (2.8,-2) {\textcolor{blue}{$O( \log m)$}};
		\draw [->] (rectitem.west) to [out=120, in = -180] (rectbundle.west);
		\node at (-2.5,-2) {\textcolor{blue}{$O( \log B)$}~\cite{cheung2008approximation}};
		\draw [->] (rectxos) to  (rectbundle);
		\node at (0.25,-0.5) {\textbf{?}};
		\draw [->] (rectmax)  to  (rectxos);
		\draw [->] (rectitem)  to  (rectmax);
		\draw [->] (rectubundle)  to  (rectmax);		
		\node at (1,-2.5) {\textcolor{red}{$\Omega( \log m)$}};
		\node at (-2,-4.5) {\textcolor{red}{$\Omega( \log m)$}};
		\node at (2,-4.5) {\textcolor{red}{$\Omega( \log m)$}};
		
		\end{tikzpicture}
	}
	\caption{Summary of the lower and upper bounds between different subclasses of pricing functions. The red font show results in this paper; blue font shows known results.}
	\label{fig:summary}
\end{figure}
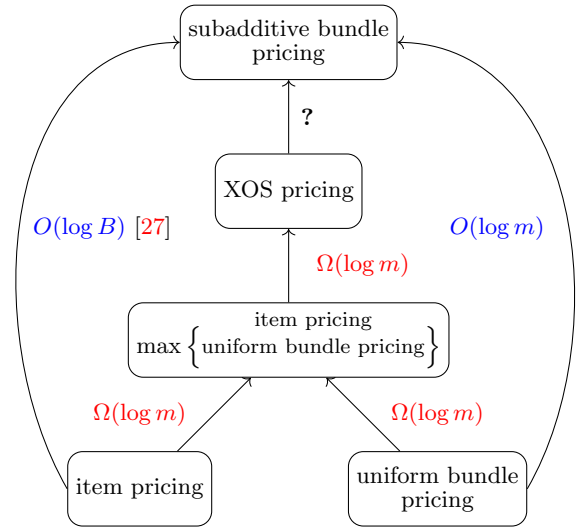

In this section, we present worst-case guarantees on how well uniform bundle pricing and item pricing can approximate the 
optimal subadditive and monotone bundle pricing. Figure~\ref{fig:summary} summarizes the upper and lower bounds that are either known, or we obtain in this paper.

\introparagraph{Upper Bounds}
It is a folklore result that for any hypergraph $\mH = (\mV, \mE)$ with valuations $\{v_e\}_{e \in \mE}$, one can always construct a uniform bundle price that is $O(\log m)$ away from
the sum of valuations $\sum_e v_e$, which is an upper bound on the optimal subadditive and monotone bundle pricing.

\begin{restatable}{lemma}{lemitembasic}
	Consider a hypergraph $\mH = (\mV, \mE)$ with valuations $\{v_e\}_{e \in \mE}$. Then, there exists a uniform bundle price $p^{b}$ that achieves revenue $O(\log m)$ away from  $\sum_{e \in \mE} v_e$, where $m = |\mE|$.
\end{restatable}

 Similarly, we know from~\cite{cheung2008approximation} that item pricing can achieve a $O(\log B)$ approximation of the sum of valuations. Recall that 
$B$ is the maximum number of hyperedges that any vertex can be contained in, and hence $B \leq m$. 

\introparagraph{Lower Bounds}
The theoretical upper bound of $O(\log m)$ is tight in the worst case for both uniform item pricing and bundle pricing. In particular, we show the following results (proofs in Appendix~\ref{sec:appendix}).
%

\begin{restatable}{lemma}{lemuniform} \label{lem:lb1}
	There exists a hypergraph $\mH = (\mV, \mE)$ with additive valuations, such that any uniform bundle price produces revenue $\Omega(\log m)$ from the optimal revenue
	$\sum_{e \in \mE} v_e$.
\end{restatable}

\begin{restatable}{lemma}{lemitem} \label{lem:lb2}
	There exists a hypergraph $\mH = (\mV, \mE)$ with uniform valuations, such that any item pricing solution produces revenue $\Omega(\log m)$ from the optimal revenue
	$\sum_{e \in \mE} v_e$.
\end{restatable}

\begin{restatable}{lemma}{lemmax} \label{lem:lb3}
	There exists a hypergraph $\mH = (\mV, \mE)$ with submodular valuations, such that any uniform bundle pricing and any item pricing produces revenue $\Omega(\log m)$ from the optimal revenue
	$\sum_{e \in \mE} v_e$.
\end{restatable}	

Note that for each of the above result, there exists a subadditive pricing function that can extract the full revenue. The above lower bounds tell us that there are problem instances where uniform bundle pricing will be optimal, but item pricing will behave poorly, and vice versa. Moreover, there are instances where both subclasses of pricing functions will not perform well with respect to the optimal submodular monotone function 
(which is a subset of subadditive and monotone bundle pricing).
 A straightforward corollary of the lower bound of Lemma~\ref{lem:lb3} is that even an XOS pricing function that combines a constant number of item pricing functions suffers from the 
 $\Omega(\log m)$ revenue gap. An open question here is whether an XOS pricing that uses a non-constant (but still small enough) number of item pricings can obtain a better approximation guarantee with respect to the optimal subadditive and monotone bundle pricing.

%% file: approxalgo.tex
\section{Approximation algorithms}
\label{section-approxalgo}

In this section, we present the various approximation algorithms that we consider in our experimental evaluation. We consider algorithms from two subclasses of subadditive and monotone pricing schemes: $(i)$ uniform bundle pricing, and $(ii)$ item (additive) pricing.

\subsection{Uniform Bundle Pricing} 
\label{subsection-uniformbundle}

In uniform bundle pricing, the algorithm sells every hyperedge at a fixed price $P$. Then, if the buyer has valuation $v_e \geq P$, the hyperedge (and thus, the query bundle corresponding to the hyperedge) can be sold. 
To compute the optimal uniform bundle price $P$, we use a folklore algorithm that we call \ubp . The algorithm first sorts the hyperedges in $\mE$
in decreasing order of valuation. Then, it makes a linear pass over the ordered valuations, and for every hyperedge $e \in E$ computes the revenue $R_e$ obtained
if we set the price $P = v_e$. In the end, it outputs $\max_e \{R_e\}$. It is easy to see that the algorithm runs in time $O(m \log m)$, and that it achieves an approximation
ratio of $O(\log m)$.

Uniform bundle pricing is very attractive as a practical pricing scheme, since it has a single parameter (and thus it has a concise representation), and computing the price of a new query is a trivial task. 
However, because it is insensitive to the structure of the bundle (and hence the query), it will perform very poorly when the valuations have a large difference across the hyperedges. 
We should remark here that we expect this to be the case in real-world scenarios: for instance, consider a large table and two queries: one that returns the whole dataset, and one that returns only a single row of the table. Then, it is reasonable to expect that the valuation for these two queries will generally differ by a large margin.

\begin{figure*}
	\hspace*{-2mm}
	\begin{subfigure}[t]{0.25\linewidth} 
		\includegraphics[scale=0.29]{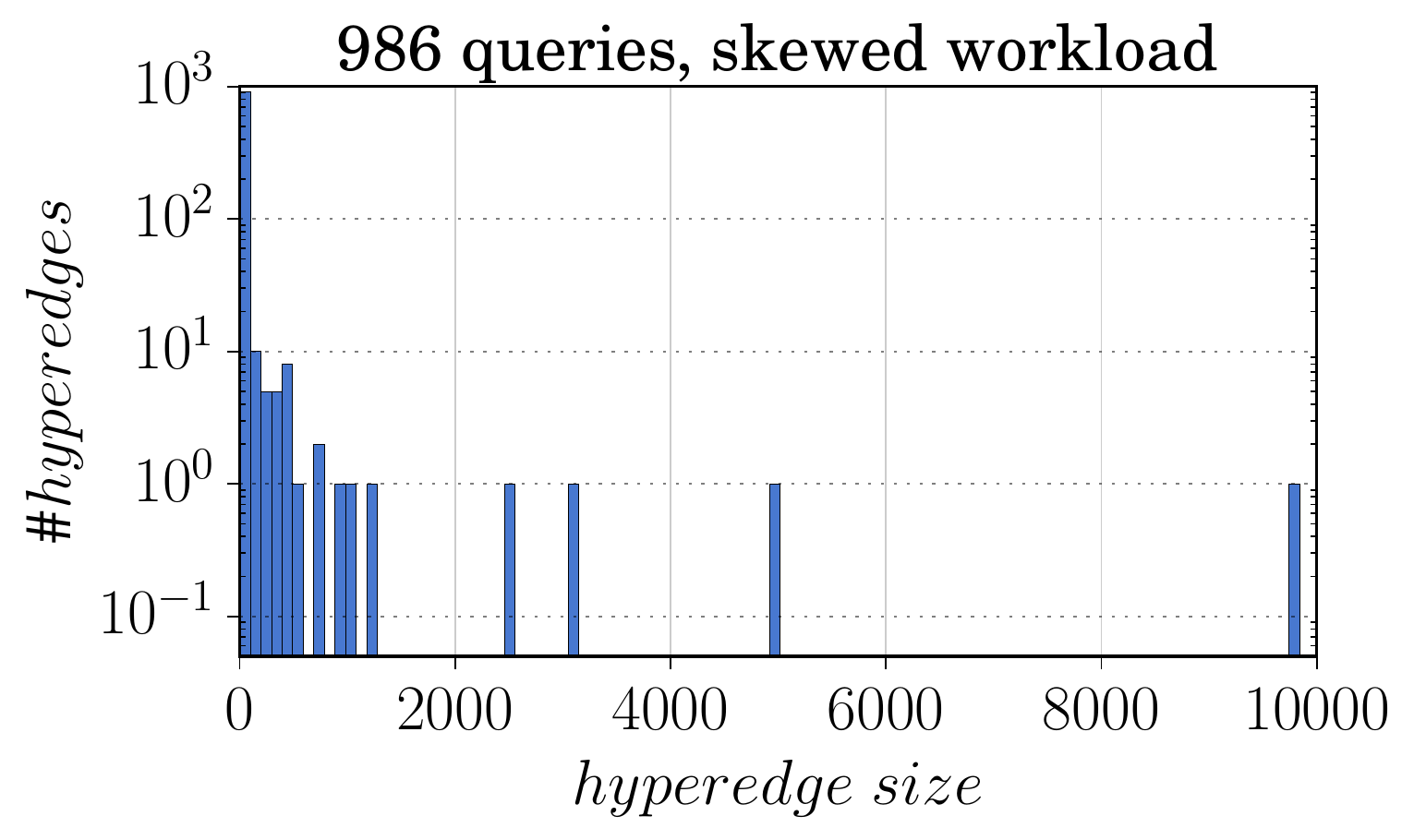}
		\caption{Skewed  workload} \label{fig:histogramrealqueries}
	\end{subfigure}       	\hspace{-2mm}
	\begin{subfigure}[t]{0.24\linewidth}
		\includegraphics[scale=0.29]{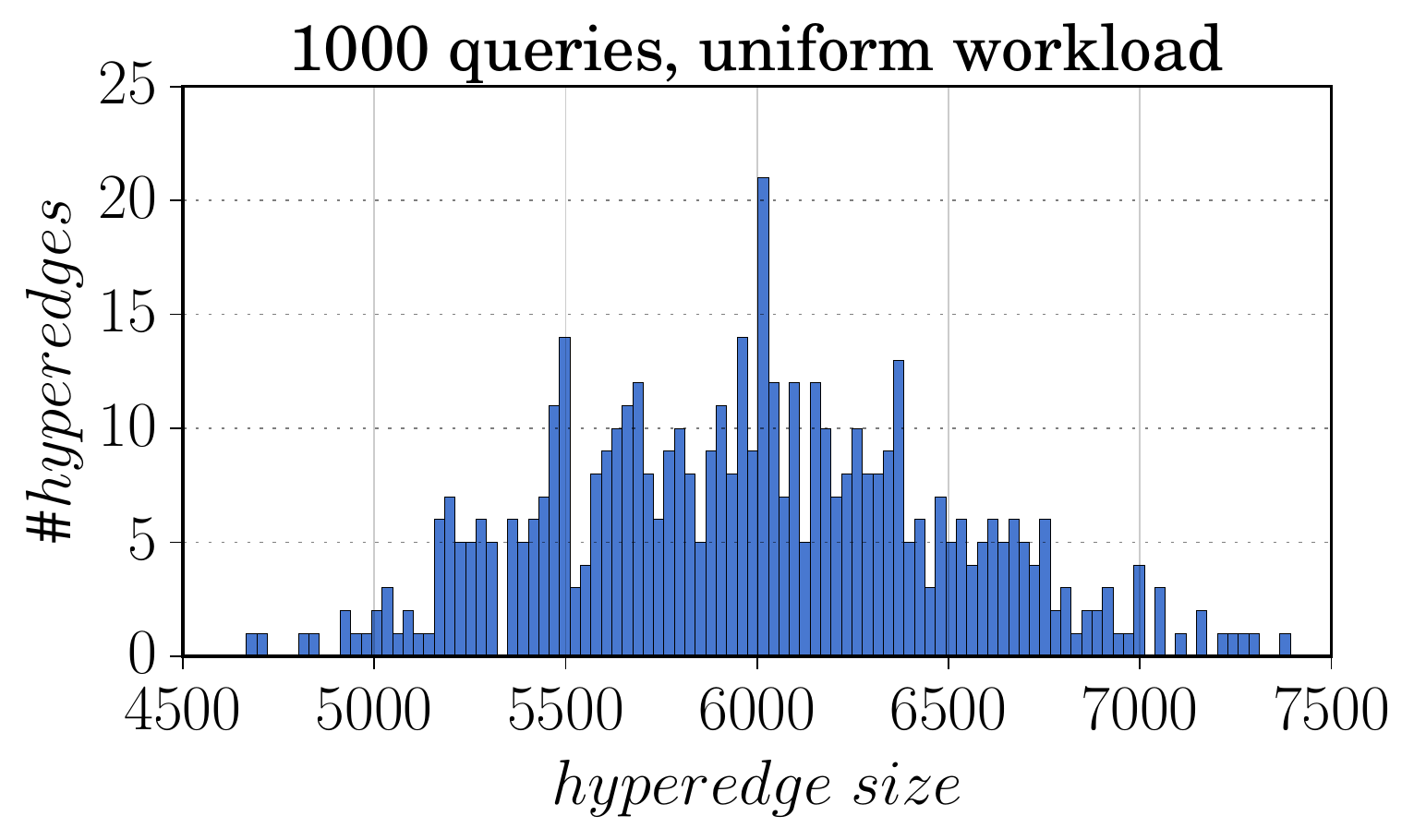}
		\caption{Uniform  workload} \label{fig:histogramrandom}
	\end{subfigure} 
	\begin{subfigure}[t]{0.24\linewidth}
		\includegraphics[scale=0.29]{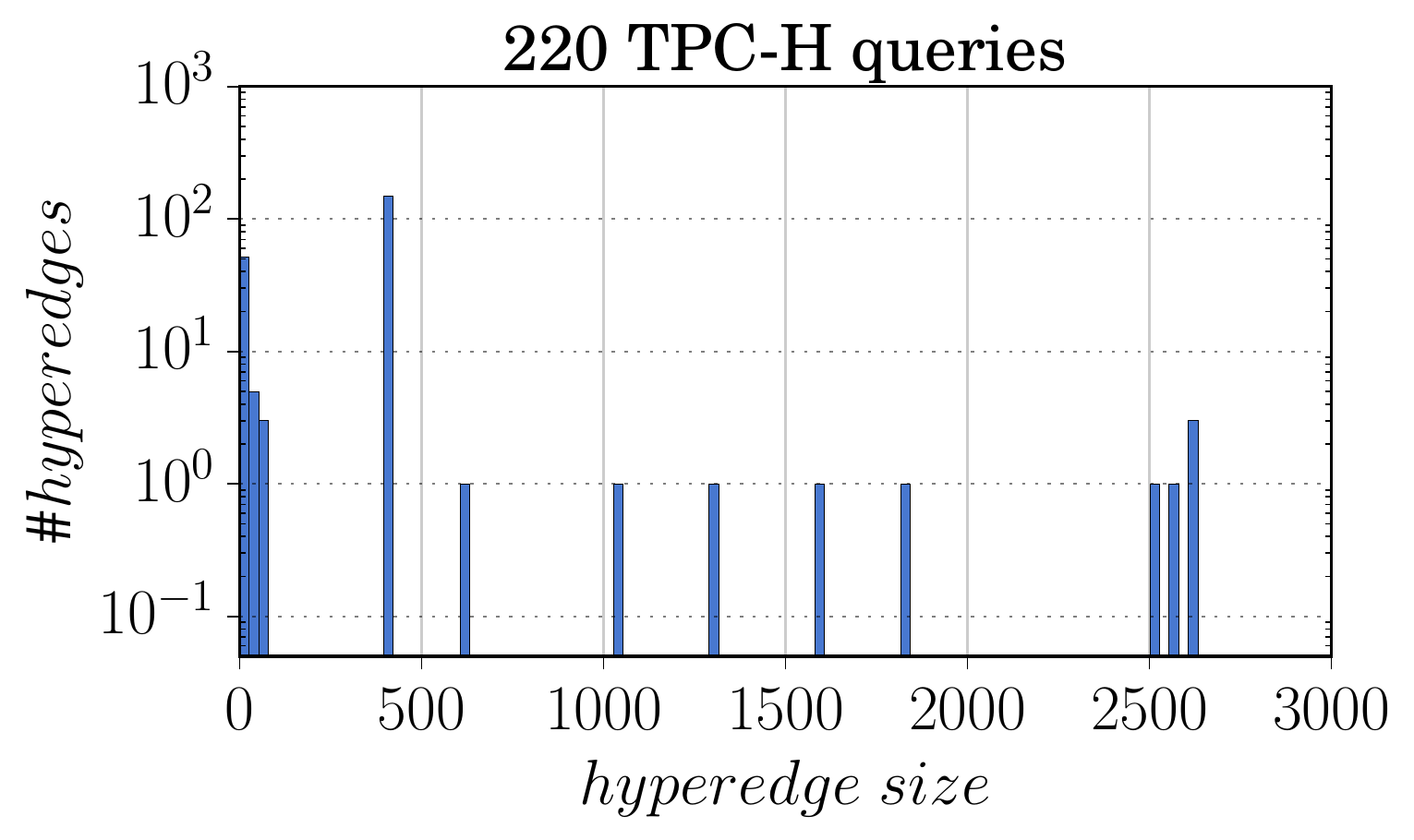}
		\caption{TPC-H  workload} \label{fig:histogramtpch}
	\end{subfigure} 
	\begin{subfigure}[t]{0.24\linewidth}
		\includegraphics[scale=0.29]{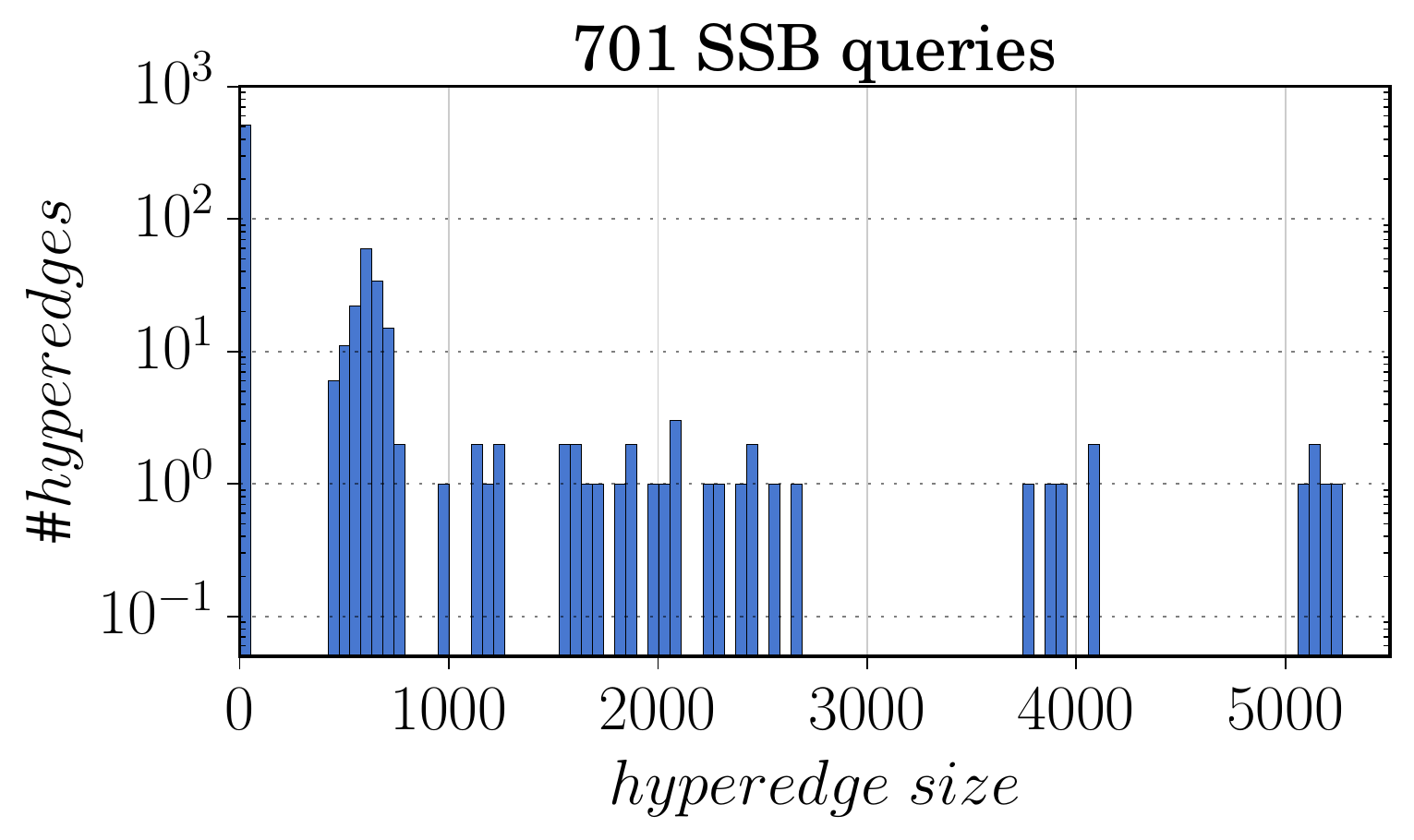}
		\caption{SSB  workload} \label{fig:histogramssb}
	\end{subfigure} 
	\caption{Hyperedge size distribution}
\end{figure*}  

\subsection{Item Pricing} 

In item pricing, we seek to assign a weight $w_j \geq 0$ to each vertex in the hypergraph. 
Then, the price of any hyperedge $e$ is given by $p(e) = \sum_{j \in e} w_j$. The representation size of item pricing is $O(n)$, so we can guarantee that it will have a 
concise representation as long as we pick the support set to be small enough (recall that $n = |\mS|$). Unlike uniform bundle pricing, item pricing can capture large differences between
the valuation of different queries. On the other hand, we also need to choose a large enough support size, so that we can extract a reasonably good revenue from our pricing
function. A large support size also guarantees that that new queries that arrive will have
non-empty hyperedges and hence will not be priced to 0. In our experiments, we evaluate four item pricing algorithms.

\smallskip
\introparagraph{The Uniform Item Pricing (\uip) Algorithm}
Our first item pricing algorithm is an $O(\log n + \log m)$-approximation algorithm given by Guruswami et. al~\cite{guruswami2005profit}. \uip\ outputs a uniform item pricing,
where every $w_j$ is set to the same value $w$. The algorithm sorts all hyperedges on the value $q_e = \frac{v_e}{|e|}$, computes for every hyperedge $e$  
the revenue $R_e$ that we can obtain if we set $w_j = q_e$ for every $j \in \mV$, and finally outputs the pricing that gives $\max_e \{ R_e\}$. Its running time is also $O(m \log m)$.

\smallskip
\introparagraph{The LP Item Pricing (\lpip) Algorithm}
The second item pricing algorithm we consider builds upon the \uip\ algorithm to construct a non-uniform item pricing. \lpip\ constructs a separate linear program $LP(e)$ for every hyperedge $e \in \mE$ as follows. Let $F_e \subseteq \mE$ be the set of hyperedges $e'$ such that $v_{e'} \geq v_e$. Then, $LP(e)$ has the objective of maximizing the revenue,
with the constraint that every edge in $F_e$ must be sold: in other words, for every $e' \in F_e$, we must have $\sum_{j \in e'} w_j \leq v_{e'}$. Observe that the uniform item pricing
solution that sets each weight $w_j$ to $v_e$ is a feasible solution for $LP(e)$, so the output of the linear program can only give a better item pricing. \lpip\ outputs the revenue-maximizing solution across all $LP(e)$. The worst-case approximation guarantee of \lpip\ is also $O(\log m)$; however, as we will see in the experimental section, it often outperforms \uip.

\smallskip
\introparagraph{The Capacity Item Pricing (\cip) Algorithm}
This algorithm is an $O(\log B)$-approximation algorithm given by ~\cite{cheung2008approximation}.
Although this primal-dual algorithm was presented in the context of item pricing with limited supply, it readily extends to the unlimited supply setting. Intuitively, \cip\ sets a uniform capacity constraint $k$ on how many times each item (vertex) can be sold, and for that $k$ solves a linear program that optimizes for the welfare-maximization problem.  The dual solution of this LP gives the prices of items such that at least $k$ copies of each item are sold. \cite{cheung2008approximation} proves that if we search through the possible capacity constraint using a step-size of $(1+\epsilon)$ -- so $k = 1, (1+\epsilon), (1+\epsilon)^2, \dots$ -- then the revenue-maximizing item pricing across all $k$'s achieves an approximation ratio of $O((1+\epsilon) \log B)$.

%
%
\smallskip
\introparagraph{The Layering Algorithm}
Since the previous algorithms require solving multiple linear programs, they can be slow when the size of input is large.  We thus also
consider a fast greedy algorithm that achieves an $O(B)$-approximation in the worst case (but as we will see, a much better approximation
in practice).

\begin{algorithm}[!htp]
	\SetCommentSty{textsf}
	\DontPrintSemicolon 
	\SetKwFunction{proc}{\textsf{eval}}
	\SetKw{KwGoTo}{go to}	
	\SetKwData{maxsum}{maxrev}
	\SetKwData{fixedprice}{fixedprice}
	\SetKwData{temp}{temp}	
	\SetKwInOut{Input}{\textsc{input}}\SetKwInOut{Output}{\textsc{output}}
	\Input{Hypergraph $\mH = (\mV, \mE)$ and valuation $\{ v_e\}_{e \in \mE}$}
	\Output{Item pricing $w_j$ for each item $j \in \mV$}
	\BlankLine
	$Rev \leftarrow 0$, $S\leftarrow \emptyset$, $w_j \leftarrow 0$ for each $j \in \mV$;\\
	\While{$\mE \neq \emptyset$}{
		Let $\mE'\subseteq \mE$ be a minimal set cover of the items in $\bigcup_{e \in \mE} e$;\\
		\If{$\sum_{e\in \mE'}v_e>Rev$}{
			$S\leftarrow \mE'$;\\
			$Rev\leftarrow \sum_{e\in \mE'}v_e$;\\
		}
		$\mE\leftarrow \mE \setminus \mE'$;
	}
	\For{$e\in S$}{
		Find item $j \in e$ such that $j \not\in e'$, $\forall e'\in S$, $e'\neq e$;\\
		$w_j \leftarrow v_e$; 
	}
	\KwRet{$\{w_j\}_{j \in \mV}$}
	\caption{The Layering Algorithm}
	\label{algo:layering}
\end{algorithm}

The key idea of the algorithm is to arrange the hyperedges in a layered fashion such that in each layer, every hyperedge has a unique item. Then, setting the weight for unique items to the valuation of the edge and all other items to zero can extract the full revenue in a particular layer. The following theorem proves the correctness of Algorithm \ref{algo:layering}, and analyzes its performance.

\begin{restatable}{theorem}{bapprox}
\label{thm-Bapprox}
Algorithm~\ref{algo:layering} outputs a $B$-approximation item pricing in $O(Bm)$ time. 
\end{restatable}

\begin{proof} \label{thm:lb1}
	Each step the algorithm finds a minimal set cover of the remaining items, call the set cover a \textit{layer}.
	On one hand, since each item presents in at most $B$ hyperedges, there are at most $B$ layers as at each step the degree of each item decreases by at least 1. On the other hand, each hyperedge $e$ in a minimal
	set cover $\mE'$ must contain at least one unique item that is not contained in other sets: otherwise $\mE'\setminus \{e\}$ is still a set cover,
	which contradicts the minimality of $\mE'$. Pricing these unique items at price equal to the value of corresponding sets can extract full revenue from the hyperedges in this layer. There must exist a layer such that the item pricing can achieve 
	$\Omega(\frac{1}{B})$ of the total value of all the hyperedges, thus this item pricing algorithm has approximation ratio $O(B)$. The running time for each step is $O(m)$, and the total running time is $O(Bm)$.
\end{proof}

\introparagraph{The XOS Pricing (\xos) Algorithm}
The last pricing algorithm we consider is the \xos\ function obtained by computing the bundle price using pricing vector from \lpip\ , \cip\ and then using the higher of the two as  price offered by the seller.

%% file: experiments.tex
\section{Experimental Evaluation}
\label{sec:experiments}

In this section, we empirically evaluate the performance of the five pricing algorithms presented 
in Section~\ref{section-approxalgo}. We evaluate the performance across two measures: $(i)$ the runtime of
the algorithm, and $(ii)$ the revenue that the algorithm can generate. 
All pricing algorithms run on hypergraph structures that are generated from a workload of
\texttt{SQL} queries executed over a real-world dataset. The valuations are obtained using different generative random processes, so as to observe the algorithmic behavior under different scenarios. These generative models are motivated by studies modeling valuations for digital goods in online platforms and their pricing~\cite{naldi2008performance, bhattacharjee2003economic, hartline2008optimal, shiller2011music, zheng2017online, bakos1999bundling, zheng2017trading}.


\subsection{Experimental setup}

\begin{figure*}
	\captionsetup[subfigure]{oneside}
	\hspace{-1mm}
	\subcaptionbox{Sampling bundle valuations \label{fig:unifzipfian}}%
	[.49\linewidth]{\includegraphics[scale=0.265]{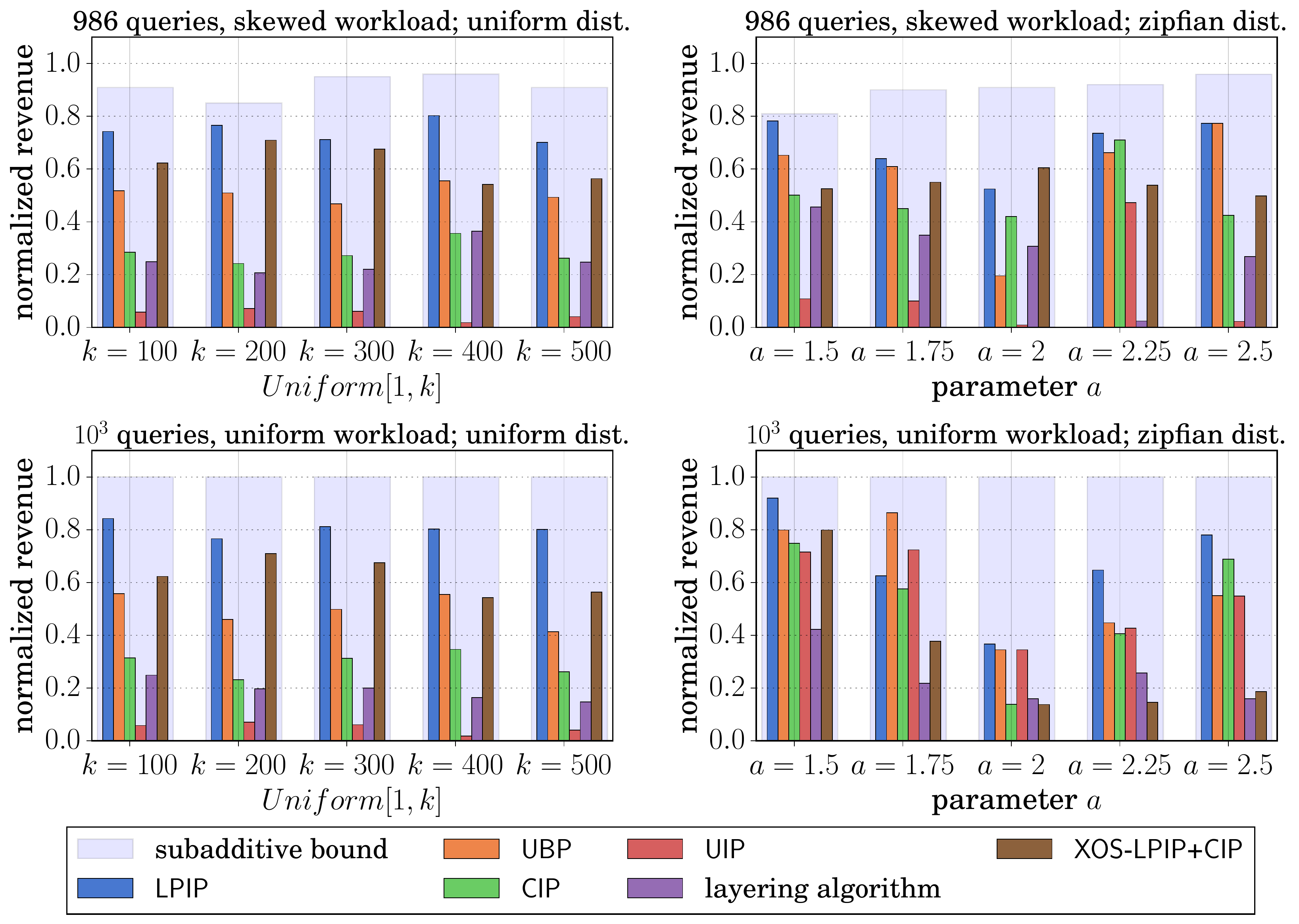}}
	\subcaptionbox{Scaling bundle valuations \label{fig:scalingedge}}%
	[.49\linewidth]{\includegraphics[scale=0.265]{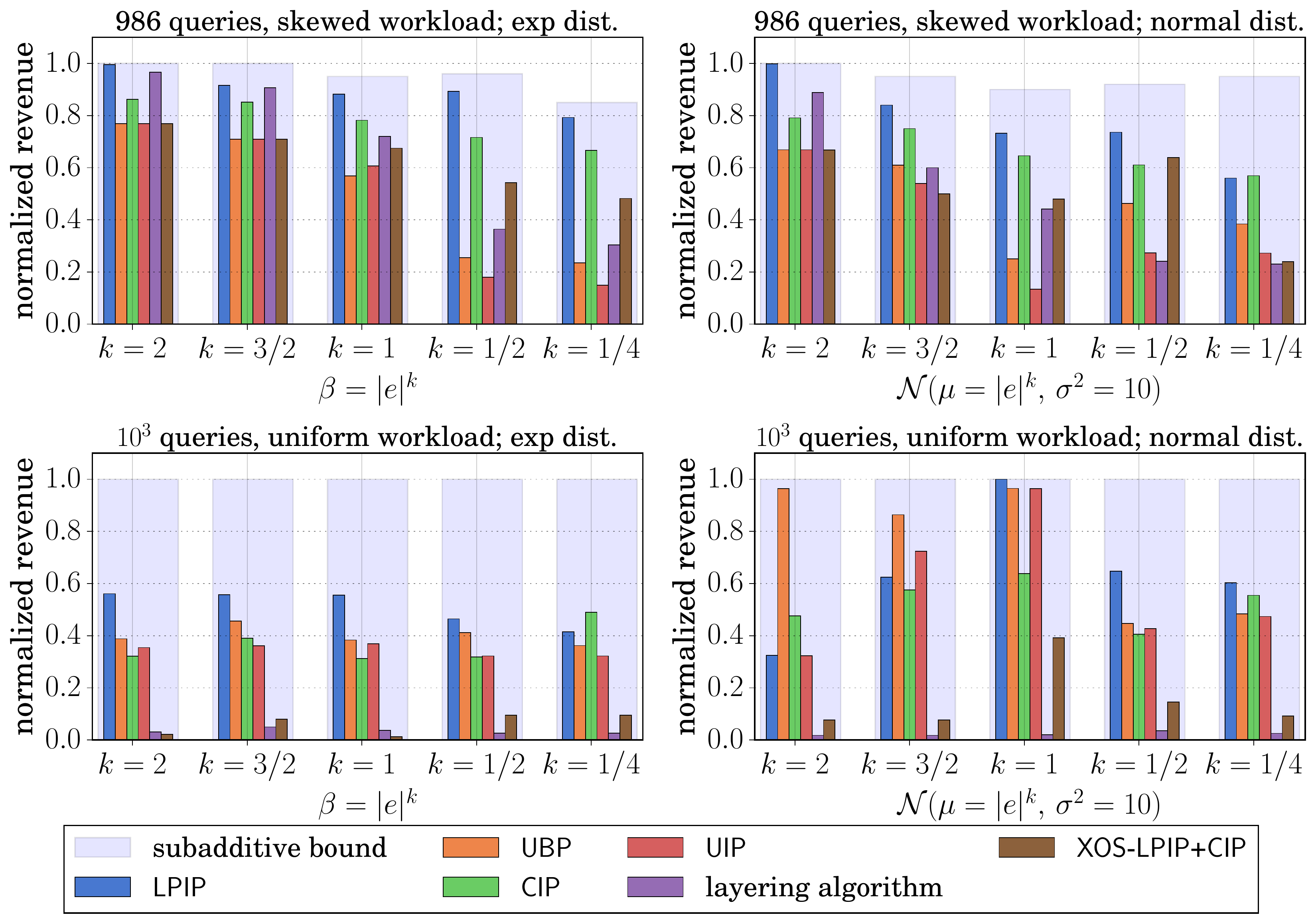}}
	\caption{skewed and uniform workload}	
\end{figure*}  

We perform all our experiments on Intel Core i$7$ processor machine and $16$ GB main memory running OS X $10.10.5$. We use \texttt{MySQL} as the underlying database for query processing and evaluation. Our implementation is written in \textsf{Python} on top of the \textsc{Qirana} query pricing system~\cite{deep2017qirana}. For all our experiments, we use \textsf{Cvxpy}~\cite{cvxpy} optimization toolkit for running linear programs. 
\textsc{Qirana} generates a support set $\mS$ by randomly sampling "neighboring" databases of the
underlying database $\db$, \ie databases from $\mI$ that differ from $\db$ only in a few places.
The advantage of this strategy is that it is possible to succinctly represent the support set by 
storing only the differences from $\db$, which is efficient in terms of storage.
For every query bundle $\bQ$, \textsc{Qirana} computes the conflict set $\dagr{\mS}{\bQ,\db}$, which
is the bundle (or hyperedge) that we use as input to the pricing algorithms.

Table~\ref{table:experiments} shows the design space of our experimental evaluation. 
Our experiments are over the $\texttt{\bfseries world}$ dataset, \tpch\ benchmark and \ssb\ benchmark. Each query workload will generate a 
hypergraph; to assign valuations over the hyperedges, we sample from different types of distributions, 
which we describe later in detail. We evaluate our algorithms for each instance that is
generated in this fashion. 

In order to compare how well our algorithms perform in terms of revenue, we use two upper 
bounds: $(i)$ sum of valuations, and $(ii)$ an upper bound on the optimal subadditive valuation. We find an upper bound on the optimal subadditive valuations by computing a linear program whose constraints encode the arbitrage constraints. Since the number of constraints can be exponential in the number of hyperedges, we optimize by greedily adding constraints for bundles with largest valuations and finding a set of bundles that cover the hyperedge with small valuations. As we will see later, this helps us compare the performance of algorithms with respect to the subadditive bound, which can
generally be much smaller than the sum of valuations. 
In all our experiments, we report each data point as an average over $5$ runs, where we discard the first run to minimize cache latency effects on running time of the algorithms.

\begin{table} \centering
	\caption{Experimental Design Space}
	\hspace*{-2.2mm}
	\scalebox{0.80}{
		\begin{small}
			\def\arraystretch{1.1}%
			\begin{tabular}{c | c | c | c}
				\toprule
				\textbf{Dataset} & \textbf{Algorithms} & \textbf{Query Workload} & \textbf{Valuation Model}\\ \toprule
				\texttt{\bfseries world} dataset & \ubp & uniform & sampled bundle  \\ \cmidrule{3-3}
				\cmidrule{1-1}& \uip & skewed & scaled bundle  \\ \cmidrule{3-3}
				\ssb\ benchmark& \lpip & \ssb\ queries & additive bundle  \\ \cmidrule{3-3}
				\cmidrule{1-1}& \cip & \tpch\ queries &  \\
				\tpch\ benchmark& Layering &  &  \\
				\bottomrule
			\end{tabular}
		\end{small}}
		\label{table:experiments}
\end{table}

\subsection{Workload and Dataset Characteristics}

\begin{figure*}
	\captionsetup[subfigure]{oneside}
	\hspace{-1mm}
	\subcaptionbox{Sampling bundle valuations: \ssb\ and \tpch\ \label{fig:unifzipfian:ssbtpch}}%
	[.49\linewidth]{\includegraphics[scale=0.26]{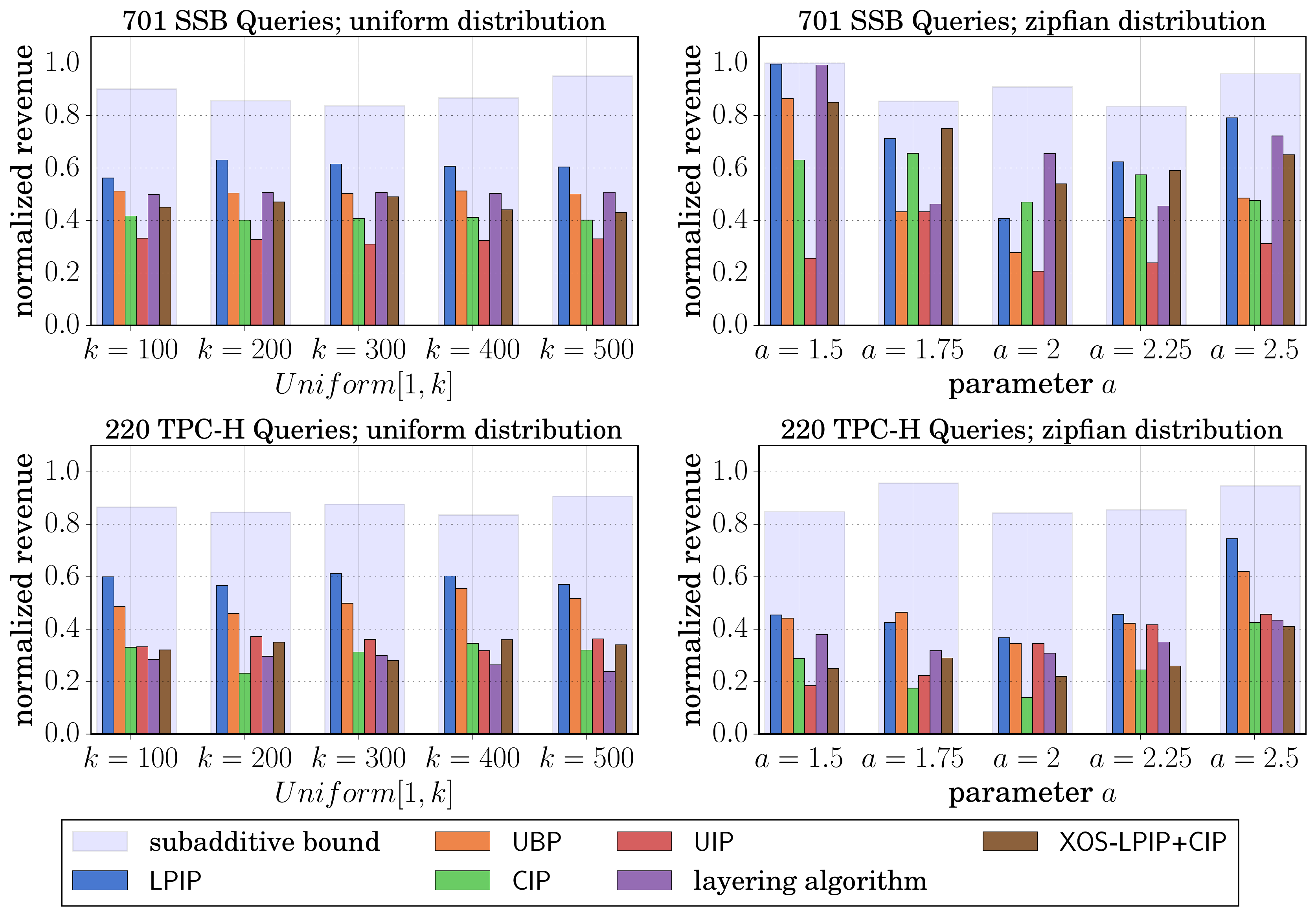}}
	\subcaptionbox{Scaling bundle valuations: \ssb\ and \tpch\ \label{fig:scalingedge:ssbtpch}}%
	[.49\linewidth]{\includegraphics[scale=0.26]{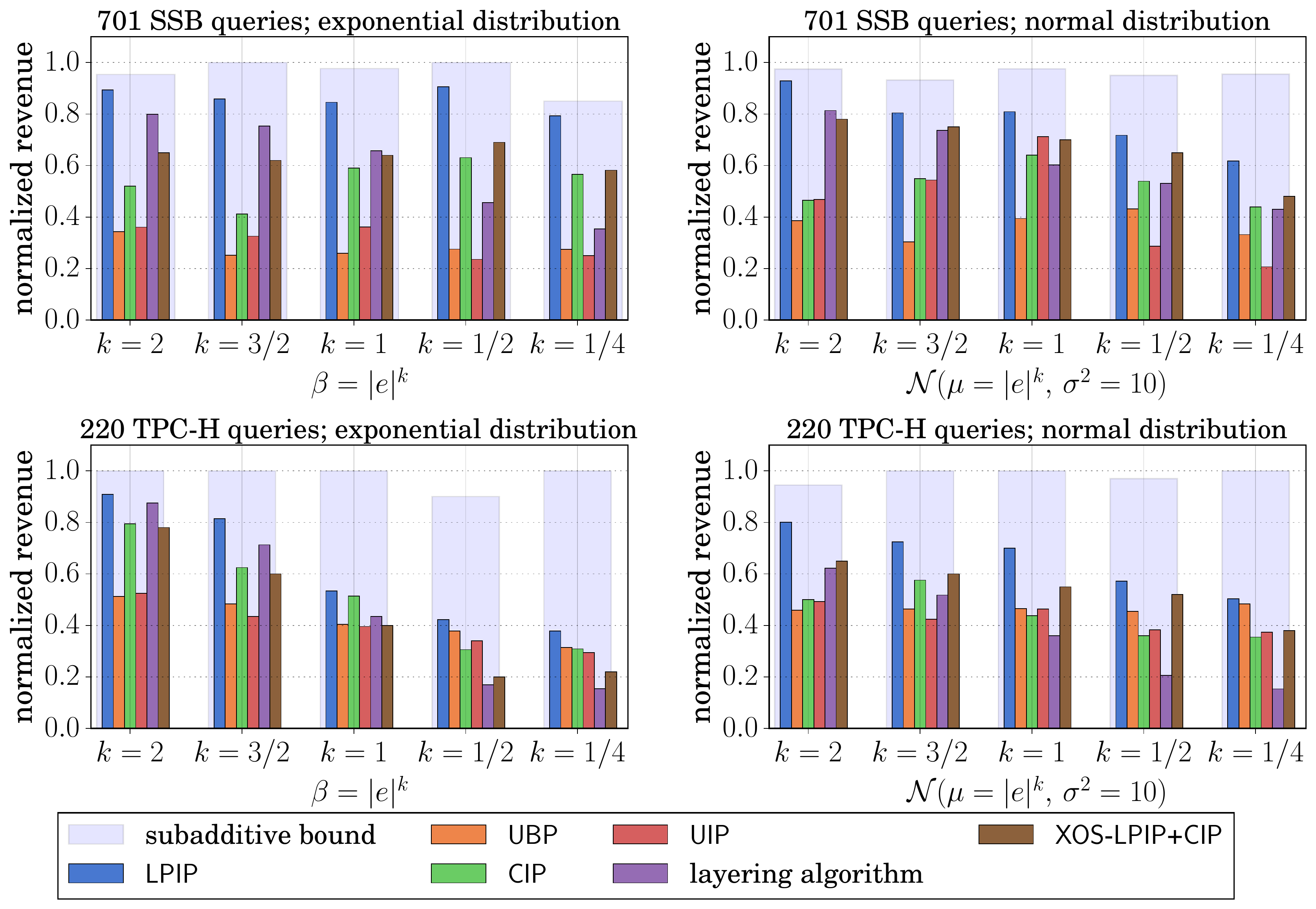}}
	\caption{\ssb\ and \tpch\ workload}
\end{figure*}  

We now describe briefly the characteristics of the query workload and datasets. The first dataset we consider is $\texttt{\bfseries world}$ dataset, a popular database provided for software developers. 
It consists of $3$ tables, which contain $5000$ tuples and $21$ attributes. We construct a support set of size $n = |\mS| = 15000$. For \tpch\ and \ssb, we generate data for scale factor of one ($\approx 10$ million rows) and support set of size $100000$.

We consider four different query workloads, which create different hypergraphs that fundamentally differ in structure:
\begin{itemize}
\item The {\em skewed} query workload~\cite{henglein2010generic} consists of $m = 986$ SQL queries containing selection, projections and join queries with aggregation. The list of queries in this workload is presented in the Appendix~\ref{sec:appendix}. 

\item The {\em uniform} query workload consists of only selection and projection SQL queries with the same selectivity (which means that the output of each query is about the same). 

\item The \ssb\ query workload is generated by using the standard twelve queries as templates where we change the constants in the predicates.

\item The \tpch\ query workload is generated by using seven of the $22$ queries that are supported by~\cite{deep2017qirana} as templates where we change the constants in the predicates. The query generation process is described in Appendix~\ref{sec:tpch:ssb}.
\end{itemize}

Table~\ref{table:workload:characteristics} summarizes the characteristics of the two hypergraphs
generated by each query workload. Both hypergraphs have the same number of vertices and
hyperedges. On the other hand, their structure is very different, as can be seen in 
Figures~\ref{fig:histogramrealqueries} and~\ref{fig:histogramrandom}, which depict the distribution of the hyperedge size. For the uniform query workload, the average size of each hyperedge is around $6000$, and it is normally distributed around that value. This means that there is a high overlap among the vertices of the hyperedges. For the skewed query workload, most of the hyperedges contain only a very small number of vertices, while only a few hyperedges contain a large number of vertices. Observe also that the average hyperedge size is around $40$, so the hypergraph is more sparse compared to the uniform query workload. \highlight{\tpch\ and \ssb\ workloads also have skew in their hyperedge distribution (Figures~\ref{fig:histogramssb} and~\ref{fig:histogramtpch}). \ssb\ workload has exactly one hyperedge with size zero and has close to half of the edges with a unique item in it. \tpch\ workload has eleven edges with size zero but only a quarter of edges have a unique item in them.}

\begin{table}
	\caption{Hypergraph Characteristics}
	\hspace{-2mm}
	\scalebox{0.81}{
	\def\arraystretch{1.3}%
	\begin{small}
		\begin{tabular}{@{}lrrr@{}}\toprule
			\textbf{Query Workload} & \textbf{\# Queries ($m$)} & \textbf{Max degree ($B$)} & \textbf{Avg edge size} \\ \midrule
			\textbf{uniform} &  1000 & 400 & 5982.07   \\ \hdashline
			\textbf{skewed} &  986 & 22 & 41.67    \\ \hdashline
			\ssb\ & 701 & 257 & 278.72 \\ \hdashline
			\tpch\ & 220 & 151 & 375.48 \\ 
			\bottomrule
		\end{tabular}
	\end{small}}
	\label{table:workload:characteristics}
\end{table}

\subsection{Measuring the Revenue }

We first focus on the behavior of the pricing algorithms with respect to the goal of 
maximizing the revenue. We will examine separately the algorithmic behavior for
different structure of the valuations.

\smallskip
\introparagraph{Sampling Bundle Valuations} 
In this part of the experiment, we generate valuations for every hyperedge by sampling 
from a parametrized distribution.

First, we sample valuations from the uniform distribution $\textsf{Uniform}[1,k]$ for some parameter $k$.
Figures~\ref{fig:unifzipfian} and~\ref{fig:unifzipfian:ssbtpch}  shows the performance of all six pricing algorithms for all workloads. We should
remark that the revenue plotted is normalized with respect to the sum of valuations including
the subadditive upper bound. 
We can observe that the \lpip\ algorithm performs much better in all cases for both query
workloads. The second best algorithm is \ubp; we expect that uniform bundle pricing performs
well in this case, since the size of the bundle is not correlated with the valuation in this setup.
Finally, notice the huge gap between \uip\ and \lpip: this is an instance where both algorithms have
the same worst-case guarantees, but their revenue differs by a large margin. \highlight{For \ssb\ and \tpch\ workloads, layering algorithm gets close to half and a quarter of the possible revenue for uniform bundle valuations (figure~\ref{fig:unifzipfian:ssbtpch}), in proportion to the number of edges with unique items. For all workloads, \xos\ pricing function obtained using the \lpip\ and  \cip\ pricing vector is close to the performance of \cip.}

Second, we sample valuations from the zipfian distribution \\ parametrized by $a$.
\lpip\ again performs better than the other pricing algorithms, but now \ubp\ comes a close second
(and in one case performs better than \lpip).  

The layering algorithm does not perform well except in the case of zipfian distribution with exponent smaller than $2$. Indeed, for $a < 2$, zipfian distribution assigns a large valuation to some hyperedge that contributes significantly to the total revenue. In such cases, the layering algorithm can always extract full revenue from the layer containing high valuation edges and perform well in practice. \highlight{This also explains why for \ssb\ workload and $a=1.5$ (figure~\ref{fig:unifzipfian:ssbtpch}), the revenue extracted is close to $1$. In that specific instance, one edge was assigned a valuation of $1428920$ and the sum of all valuations was $1653537$}. As the zipfian exponent becomes greater than two, the spread of valuations becomes smaller and the layering algorithm performs worse.

Finally, the \cip\ algorithm does not perform that well, even though it is theoretically optimal. This is because going over all capacity vectors with limited supply is very expensive. In our implementation, running the linear program for a large number of capacity vectors for the uniform workload takes close to $2$ hours in total (we discuss reasons for this in the next section). Thus, we reduce the number of capacity vectors that we try by increasing the $(1+\epsilon)$ parameter. This introduces a factor of $(1+\epsilon)$ in the approximation ratio but allows for the running time to be smaller. For the purpose of experiments, we set $(1+\epsilon)$ such that the running time is $\sim 30$ minutes. Performing this optimization allows us to complete the algorithm rather than truncating the experiment prematurely and returning the best result obtained so far.  The approximation factor of \cip\ remains marginally inferior to \lpip\ (although in some cases, it outperforms \lpip) while \xos\ pricing is consistently worse than both \lpip\ and \cip.
\begin{figure*}
	\captionsetup[subfigure]{oneside}
	\subcaptionbox{Sampling item prices: uniform and skewed workload \label{fig:mixing}}%
	[.49\linewidth]{\includegraphics[scale=0.26]{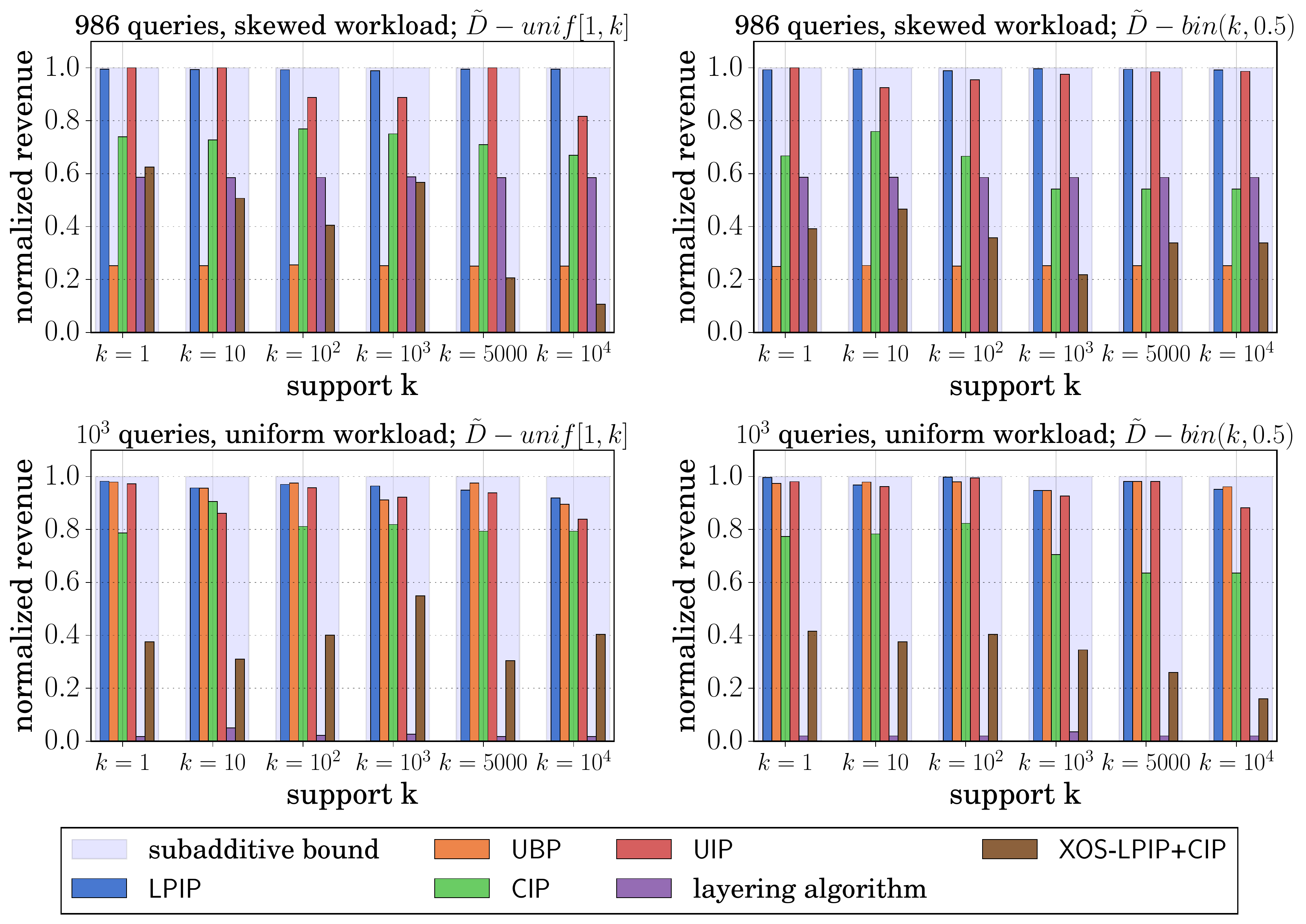}}
	\subcaptionbox{Sampling item prices: \ssb\ and \tpch\ \label{fig:mixing:ssbtpch}}%
	[.49\linewidth]{\includegraphics[scale=0.26]{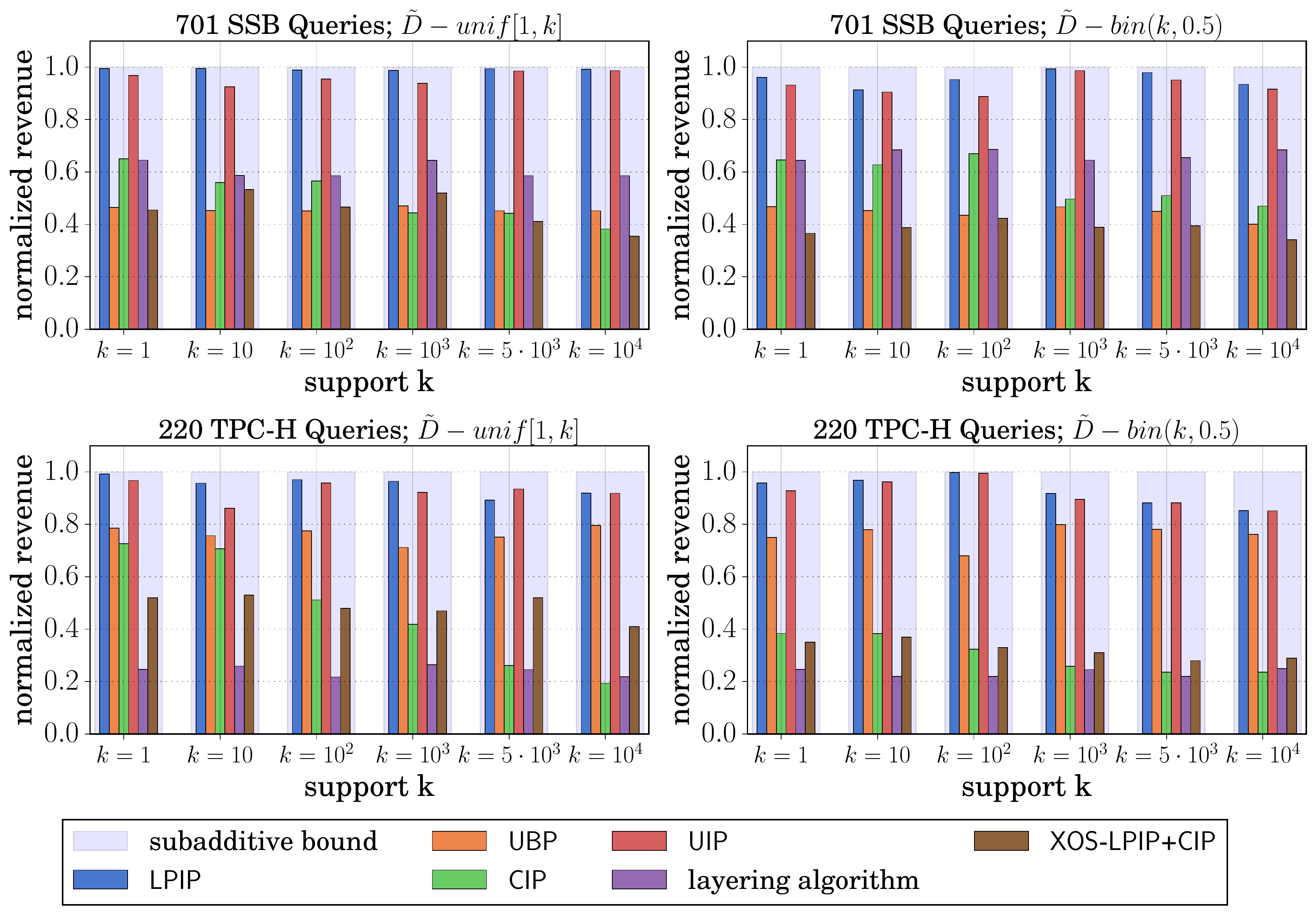}}
	\caption{Sampling item prices: all workloads}
\end{figure*}  

\smallskip
\introparagraph{Scaling Bundle Valuations} In the previous scenario, the valuations were sampled independently of the edge size. Our next experiment correlates the size of each hyperedge with the valuation that is assigned to it. 
To achieve this, we sample each valuation from the parameterized exponential and normal distribution as follows: we assign $v_e \sim {\rm exponential} (\beta = |e|^k)$ where $\beta$ is the mean of the distribution. Similarly, for normal distribution, we assign $v_e \sim \mathcal{N}(\mu = |e|^k,\, \sigma^2 = 10)$. Here $k$ is the parameter that we will vary. Figure~\ref{fig:scalingedge} and~\ref{fig:scalingedge:ssbtpch} shows the revenue
generated for the four query workloads and two families of distributions for 
different values of the parameter $k$. 

For the skewed query workload, when $k \geq 1$, most of the revenue is concentrated in a few edges that have extremely large valuations (because there are few edges of large size). In this case, all algorithms perform very well and extract almost all of the revenue. For smaller values of $k$, the algorithms can extract smaller revenue, and \lpip\ and \cip\ perform best. Notice again the large revenue gap between \lpip\ and \uip\ for such values (as much as 5x). \highlight{The same behavior is also observed for \tpch\ and \ssb\ . Since layering algorithm is the second best algorithm for $k > 1$, we investigated why this is the case for \tpch. We found that all hyperedges with size greater than $500$ (total of $10$ as seen in figure~\ref{fig:histogramtpch}) have unique items and thus are placed in a single layer. Since these edges contribute the most to the revenue, layering performs close to the best. \ssb\ workload has a more even spread of edge with unique items. Out of $36$ edges with size $> 1000$, $22$ edges have unique items.}

The landscape changes for the uniform query workload. The main observation is that the layering
algorithm performs extremely poorly. The revenue generated by the other four algorithms is very
close, with \lpip\ and \ubp\ performing the best.  
For the exponential distribution, all algorithms are very far from optimal. However, we believe this is not an anomaly but rather the subadditive bound not being as good as it should be.

\smallskip
\introparagraph{Sampling Item Prices} The last set of experiments is to understand the behavior of the pricing algorithms when the valuation of each hyperedge is defined by an \emph{additive generative model}. More specifically, we define $k$ different distributions $\{D_i\}_{i=1}^{k}$ from which items will draw their prices and a special distribution $\tilde{D}$ which will assign each item which distribution it will sample from. The valuation of an edge is the defined as $v_e = \sum_{j \in e} x_j \sim D_{\ell_j}$ where $\ell_j \sim \tilde{D}$. Intuitively, this model will capture the scenario where parts of the database have non-uniform value and some parts are much more valuable than others. To see why this setting is of practical interest, consider a research analyst in banking who gives stock recommendations. While public information about companies and stocks may be cheap, the research analysts buy and sell recommendations will be of much higher value. For the purpose of experiments, we fix $D_i$ to $\textsf{Uniform}[i, i+1]$ and set $\tilde{D}$ to $\textsf{Uniform}[1, k]$ or $\textsf{Binomial}(k, 1/2)$ while varying $k$. Figure~\ref{fig:mixing} and~\ref{fig:mixing:ssbtpch} shows the results of this experiment. 

Here, \lpip\ outperforms all other algorithms across all workloads. For small values of $k$, the valuation of each hyperedge is closer to $|e|$. In this case, there is no gap between \uip\ and its \lpip. As the value of $k$ increases, the gap between the two algorithms increases, since the weights of each item become less uniform.
 
We should remark two things that are distinct for each query workload. For the skewed query workload, \ubp\ performs poorly, since now the valuation of each hyperedge is correlated (in an additive fashion) with the bundle structure. For the uniform query workload on the other hand, \ubp\ does well, since the size of the edges is relatively concentrated. Finally, the layering algorithm is the worst performing out of all in the case of the uniform query workload. 

\highlight{For \ssb\ and \tpch\ , the first observation is that although these workloads are also skewed, \ubp\ performs reasonably well (often beating \cip\ and layering for \tpch). This is consistent with the fact that in the hyperedge distribution for \tpch\ , $150$ hyperedges have a size of $\sim400$. Thus, for smaller values of $k$, \ubp\ is expected to perform well. \ssb\ hyperedge distribution is more spread out compared to \tpch\ which explains why \ubp\ is not as well performing. We go one step further to see if a post-processing step can refine \ubp\ prices to boost the revenue even more. To do so, we find the best item prices via a linear program where the constraints sell all edges sold by uniform bundle price that achieves the maximum revenue for $k=1, \textsf{Uniform}[1,k]$ in \tpch\ workload. We observed that this simple step (runs in $\sim 1s$) improves the revenue from $0.78$ to $0.99$.   Layering algorithm again performs the worst for \tpch. This happens because none of the $150$ hyperedges with size $\sim 400$ contains a unique item. Thus, although valuation of each edge is correlated with $|e|$, the total revenue in the hyperedges with large size is not significantly more as compared to when the valuation is proportional to (say) $|e|^2$ as in the case of experiments in figure~\ref{fig:scalingedge:ssbtpch}. Layering performs better for \ssb\ as the edges with unique items are more evenly spread (as noted before). Thus, layering algorithm is able to extract more revenue from the large size hyperedges as compared to \tpch. Perhaps the most interesting observation is that \xos\ pricing function performs significantly worse than best of the two. This happens because the $\max$ function assigns prices to bundle that overshoots the $v_{\bQ}$ leading to lower revenue.}

\begin{table}[] 
	\caption{Algorithm running times (in seconds) for different workloads.}
	\hspace{-3mm}
	\scalebox{0.85}{
		\begin{small}
			\begin{tabular}{@{}lrrrrr@{}}\toprule
				\textbf{Query Workload} & \textbf{\lpip} & \textbf{\ubp} & \textbf{\uip} & \textbf{\cip} & \textbf{Layering}  \\ \midrule
				
				\textbf{skewed} &  60.62 & $< 1$ & 25.45 & 812.67 & 15.67 \\ \hdashline
				\textbf{uniform} &  95.81 & $< 1$ &  29.82 &1800 & 50.19 \\ \hdashline
				\ssb & $1300 + 3600$ & $< 1$ & $1300+13$ & 7200 & $1300 + 32$ \\ \hdashline
				\tpch & $2000 + 1900$& $< 1$ & $2000+13$ & 7200 & $2000+4$ \\ 
				\bottomrule
			\end{tabular}
		\end{small}
	}
	\label{table:runtime}
\end{table}

\subsection{Measuring the Runtime}

In this section, we discuss the running time of the algorithms. Table~\ref{table:runtime} shows the runtime of all algorithms \footnote{We skip \xos\ in this section as it is dependent on both \lpip\ and \cip\ making it very expensive.}. The most time efficient algorithms are uniform bundle pricing, uniform item pricing and the layering algorithm. Uniform bundle pricing and uniform item pricing depend only on number of hyperedges and the number of items in the hypergraph. Thus, they are very fast to run in practice. Note that for all item pricing algorithms, we also include the time taken to compute the conflict set of the query. However, for uniform bundle pricing, we need not take that into account as it is independent of the conflict set. For skewed and uniform workload, the layering algorithm is slightly slower but comparable in performance. Note that the layering is faster on the skewed query workload as compared to the uniform query workload, since the maximum degree $B$ is much smaller. \highlight{Note that since the support set and the underlying database is much larger for  \ssb\ and \tpch\ , running time to construct conflict set is also large ($\sim1300s$ and $\sim2000s$ respectively in table~\ref{table:runtime}).}

The two slowest running algorithms are \lpip\ and \cip\ as they require running multiple linear programs. In practice, \lpip\ is faster than \cip. This is because the size of the linear program is very different. In our setting, the number of edges $220 \leq m \leq 1000$ is much smaller than the number of items $n = 15000(100000)$. \lpip\  has at most one constraint per bundle (thus, at most $m$ constraints) but \cip\ has one constraint per item ($n$ constraints in total). This dramatically influences the running time of the two algorithms. \cip\ uses $(1+\epsilon)$ as a parameter, where $\epsilon$ controls the limited supply available for each item. We adjust the value of $\epsilon$ for both workloads to ensure that the running time is at most $30$ minutes. We fix $\epsilon = 0.2$ for the skewed workload and $\epsilon = 4$ for the uniform workload based on our empirical observations. \highlight{For \tpch\ and \ssb\ experiments, \cip\ did not run to completion for values of $\epsilon \leq 0.5$ since the itemset here is much larger. In this case, we fix a value of $\epsilon = 3$ to limit the running time to a total of $2$ hours. \lpip\ still remains the best performing algorithm for uniform and zipfian distribution.}

\subsection{Impact of Support Set Size} \label{sec:varying:support}

\begin{table}
	\caption{Algorithm running times (in seconds) for skewed workload (including hypergraph construction time)}
	\hspace{-3mm}
	\scalebox{0.99}{
	\begin{small}
		\begin{tabular}{@{}lrrrrr@{}}\toprule
			\textbf{Support Set Size} & \textbf{\lpip} & \textbf{\ubp} & \textbf{\uip} & \textbf{\cip} & \textbf{Layering}  \\ \midrule
			
			$|\mS| = 100$ &  $<1$ & $<1$ & $<1$ & $<1$ & $<1$ \\ \hdashline
			$|\mS| = 500$ &  6.16 & $<1$ &  5.25 & 6.87 & 1.6 \\ \hdashline
			$|\mS| = 1000$ &  15.10 & $<1$ &  17.43 & 29.82 & 3.12 \\ \hdashline
			$|\mS| = 5000$ &  30.12 & $<1$ &  29.82 & 189.97 & 8.78 \\ \hdashline
			$|\mS| = 15000$ &  70.42 & $<1$ &  35.21 & 676.23 & 12.34 \\
			\bottomrule
		\end{tabular}
	\end{small}
	}
	\label{table:runtime:supportsetsize}
\end{table}

\begin{table}
	\caption{Algorithm running times (in seconds) for \ssb\ workload (excluding hypergraph construction time)}
	\hspace{-3mm}
	\scalebox{0.97}{
	\begin{small}
		\begin{tabular}{@{}lrrrrr@{}}\toprule
			\textbf{Support Set Size} & \textbf{\lpip} & \textbf{\ubp} & \textbf{\uip} & \textbf{\cip} & \textbf{Layering}  \\ \midrule
			
			$|\mS| = 1000$ &  180.29 & $<1$ & $<1$ & 3.55 & $<1$ \\ \hdashline
			$|\mS| = 5000$ &  363.97 & $<1$ &  2.85 & 21.43 & $<1$ \\ \hdashline
			$|\mS| = 10000$ &  709.10 & $<1$ &  5.12 & 50.66 & $<1$ \\ \hdashline
			$|\mS| = 50000$ &  2500 & $<1$ &  11.24 & 692.97 & 6.2 \\ \hdashline
			$|\mS| = 100000$ &  3600 & $<1$ &  13.21 & 7200 & 32.58 \\
			\bottomrule
		\end{tabular}
	\end{small}
	}
	\label{table:runtime:supportsetsize:ssb}
\end{table}

\begin{figure}[t]
	\subcaptionbox{skewed workload \label{fig:supportsetsize}}%
	[\linewidth]{\includegraphics[scale=0.44]{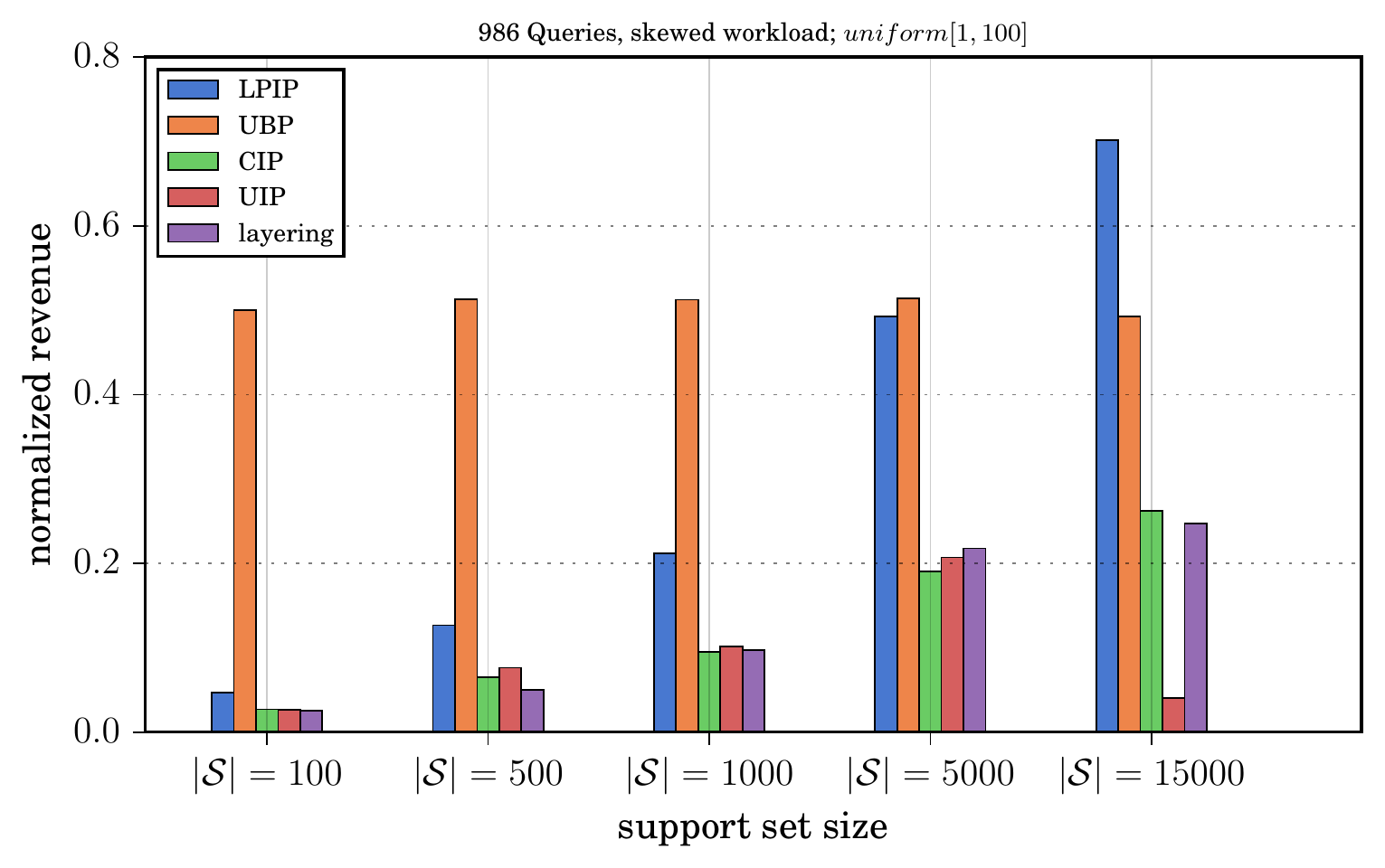} }
	\subcaptionbox{\ssb\ workload \label{fig:supportsetsize:ssb}}%
	[\linewidth]{\includegraphics[scale=0.44]{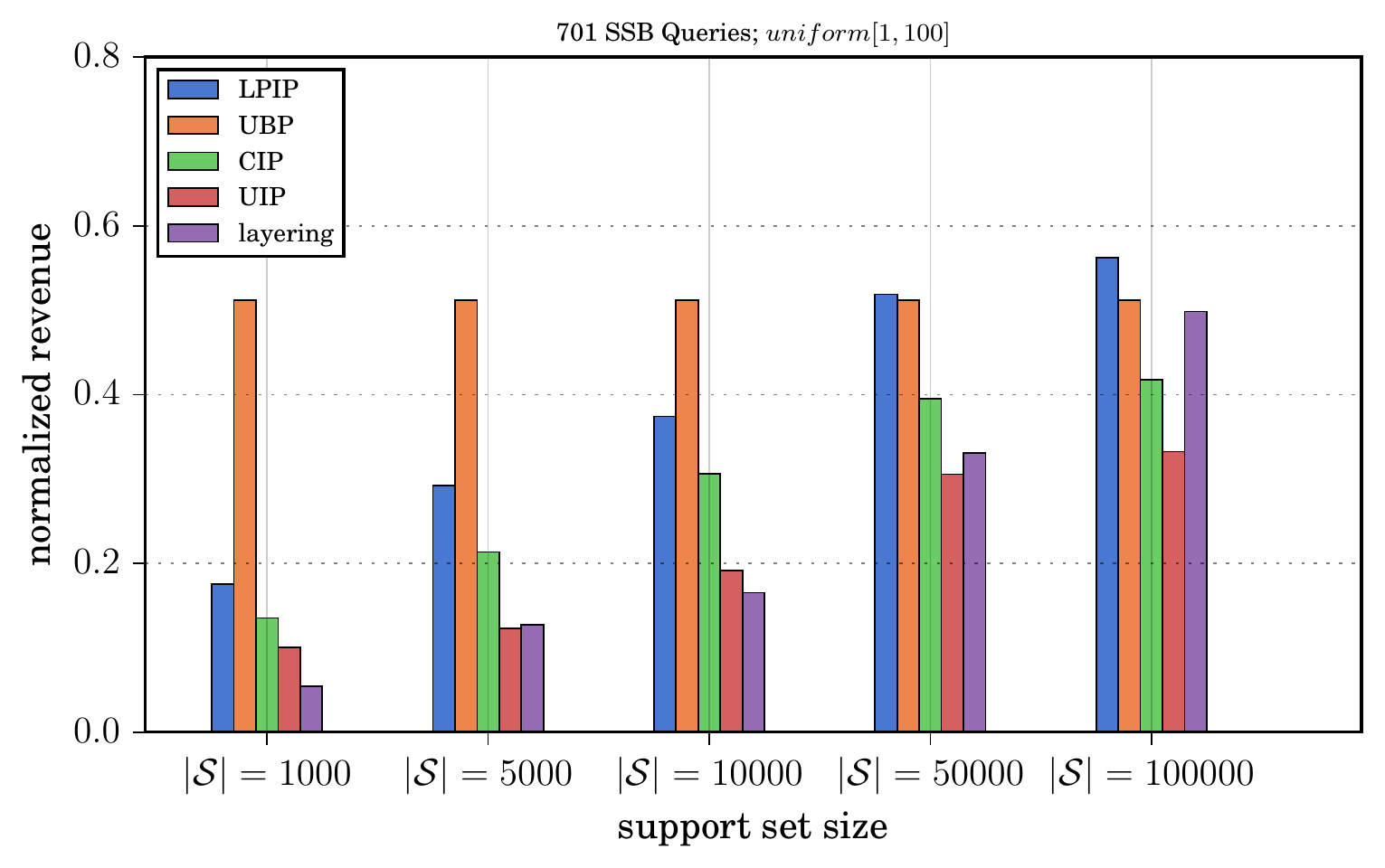} }	
	\caption{Revenue generated with increasing itemset size} 
\end{figure}

Our final set of experiments is to understand the impact of support set size, \ie, the number of items $n$. Given a hypergraph instance, adding more items to the hypergraph is an interesting proposition since it can only increase the revenue. However, this also comes at a higher cost of running time. On the other hand, too few items in the hypergraph can lead to suboptimal revenue for most algorithms (except uniform bundle pricing, since it is independent of the items). Figure~\ref{fig:supportsetsize} demonstrates the impact of changing the support set size on the revenue extracted. Unsurprisingly, uniform bundle pricing is not impacted by the support set size. As the support set size decreases for $15000$ to $100$, the  performance of all item pricing algorithms decreases. Finally, Table~\ref{table:runtime:supportsetsize} depicts the running time of all algorithms as a function of support size. The key takeaway is that running time is dependent on the support set size. Thus, the right tradeoff depends on the data seller requirements. It remains an interesting open problem to design algorithms for choosing the items in a smarter way. This will ensure that we can get good revenue guarantees without sacrificing running time. For instance, if we can create the support set in such a way that every hyperedge contains a unique item, then we can extract the full revenue from the buyers.

\highlight{Figure~\ref{fig:supportsetsize:ssb} shows the same trend in revenue drop as the support set shrinks. The running time of the algorithms is more interesting for \ssb. Observe the steep decrease in running time of  \cip\ as we go from support size $100000$ to $50000$ (table~\ref{table:runtime:supportsetsize:ssb}). This drop happens because of two reasons: $(i)$ since \cip\ contains one linear program {\em per} item, halving the support size reduces the number of linear programs by the same factor. $(ii)$ as the number of items decrease, the maximum degree of hypergraph (i.e $B$) also decreases.} 

%% file: lessons.tex
\section{Key Takeaways} \label{sec:lessons}

The empirical study has brought forward many insights. Below, we summarize some of the most important lessons from our studies and simple rules of thumb that a data broker should follow.

\subsection{Lessons Learned}

 \introparagraph{Choice of algorithm} The right choice of algorithm depends on the revenue guarantees desired by the broker, running time constraints imposed and the knowledge about query instances that need to be priced. Throughout our experiments, \cip\ has been the worst performing algorithm followed by layering algorithm (except for zipfian distribution on skewed workloads where it was either the best or second best) while \lpip\ has consistently outperformed all other choices. Thus, if the broker does not have any running time constraints, \lpip\ is the best pick. However, since \lpip\ can also be expensive (as seen for \ssb\ and \tpch), the better of layering and uniform bundle pricing gives the best empirical revenue guarantee.

\introparagraph{Hypergraph structure} Knowledge about the structure of the hypergraph is crucial in predicting the performance of the algorithm. For instance, if a non-negligible fraction of the hyperedges have size zero (TPC-H workload) or have similar edge sizes (uniform query workload), then \ubp\ performs very well. Similarly, if a large fraction of hyperedges contain a unique items, the layering algorithm and \lpip\ perform well.

 \introparagraph{Scalability challenges} Our experiments indicate a wide gap between the theoretical guarantees of the algorithms and their practical utility. Specifically, LP based algorithms suffer from scalability issues as the instance size grows. For \lpip\, there are $m$ constraints in total while \cip\ contains $n$ constraints (recall that $m \ll n$ in our setting). \lpip\ works well for small values of $m$ and $n$. As $m$ grows, \lpip\ also starts suffering from scalability issues.  On the other hand, \uip, \ubp\ and layering algorithms are both time and memory efficient.

 \introparagraph{Valuation distribution} Assumptions about valuation distribution for bundles and correlation with bundle size is also a key indicator of extracted revenue. For all query workloads, whenever the revenue is concentrated in a few edges (zipfian distribution with $a < 2$, scaled valuations and additive model), \lpip\ and \uip\ perform well. We perform experiments with different distributions and valuation models to expose the strength and weaknesses of each algorithm. However, to the best of our knowledge, there is no existing user study or datasets available that can help us validate assumptions about how buyers value different queries in the context of data pricing. 

\subsection{Future Work} Our empirical study has also given us many hints for future directions. We find the following tasks particularly urgent and engaging.

 \introparagraph{Choosing support set} Recall from Section~\ref{sec:framework} that our data pricing framework depends on a chosen set $\mathcal{I}$ of possible database instances. This choice is controlled entirely by the seller. However, a careful selection of these databases can affect the hypergraph structure. For instance, if there is a way to choose the items such that most hyperedges will have a unique item, then the pricing becomes significantly easier. More formally, we propose the following problem: Given a set of queries $Q_1, \dots Q_m$, database $\db$, does there exist a set of databases $D_1, \dots, D_m$ such that $Q_i(D_i) \neq Q_i(\db)$ but $Q_i(D_j) = Q_i(\db), i \neq j$. 
 Our goal is to study the data complexity of the problem and identify query fragments which admit efficient algorithms. A related variant of the problem is when the broker decides to fix the query templates for the buyers. Since the set of possible queries is restricted, the hypergraph structure can be controlled carefully to make pricing more amenable. 

 \introparagraph{Learning buyer valuations} This work assumes that queries and valuations are available apriori allowing for pre-processing where we can run the revenue maximizing algorithms to identify the best pricing vector. It is also worthwhile to investigate how we can learn the prices on-the-fly. In the online setting, queries arrive and the marketplace has to dynamically vary the  prices based on whether the query was bought by the buyer or not. We plan to investigate how bandit algorithms and gradient descent algorithms perform when all buyers have a fixed valuation that is unknown to the algorithm. Note that the online pricing problem requires a new model of arbitrage freeness, where the temporal aspect of the problem is also taken into account.

\introparagraph{Maximizing revenue} From a mechanism design perspective, several interesting problems remain open. First, it remains unknown what is the gap between optimal subadditive revenue and \xos pricing functions with non-constant number of additive components. The complexity of finding the optimal item prices over graphs under the additive model where each item draws its value from a distribution and the complexity of item pricing over hypergraphs with specific structure (e.g. trees) also remains open. 

 \introparagraph{User study} Finally, we believe it is worthwhile to perform a user study in order to understand how buyers interact in the data market, what queries are of interest, and how they are valued.  

%% file: conclusion.tex
\section{Conclusion}
\label{sec:conclusion}

In this paper, we study the problem of revenue maximization in the context of query-based pricing.
We cast the task as a bundle pricing problem for single-minded buyers and unlimited supply, and then perform a detailed experimental evaluation on the effectiveness of various approximation
algorithms that provide different worst-case approximation guarantees. Our results show that the specific bundle structure often means that simple item-pricing algorithms perform much better than their worst-case guarantees. There are several interesting open questions in this space. We believe that making progress on these questions is an important step, likely to create significant impact in practice.

\smallskip
\introparagraph{ Acknowledgements} Shuchi Chawla and Yifeng Teng were supported in part by NSF grants CCF$-1617505$ and CCF$-1704117$. Paraschos Koutris and Shaleen Deep were supported by funding from University of Wisconsin-Madison Office of the Vice Chancellor for Research and Graduate Education and Wisconsin Alumni Research Foundation.

%% file: appendix.tex
\section{Missing Proofs}
\label{sec:appendix}

\lemitembasic*

\begin{proof}
Consider the valuations $v_{e_1}, v_{e_2}, \dots, v_{e_m}$ in increasing order. We claim that setting $p^{b}(e)  = v_{e_j}$ for very $e \in \mE$ achieves the 
desired approximation for some edge $e_{j}$. 
The revenue $R_j$ we obtain by selling at price $v_{e_j}$ is $R_j \geq (m-j+1)v_{e_j} $. Then, the best revenue $R$ has $R \geq R_j$.
Adding up all inequalities, we get:
	
	\begin{equation*}
	\begin{aligned}
	\sum_{j=1}^m v_{e_j} \leq  \sum_{j=1}^m \frac{R_j}{m-j+1} \leq R \sum_{j=1}^m \frac{1}{j} = R \cdot O(\log m) .
	\end{aligned}
	\end{equation*}
\end{proof}

\lemuniform*

\begin{proof}
	Consider $n$ items and $m$ buyers ($n=m$), such that buyer $b_i$ wants item $i$ for price $1/i$. It is easy to see that the valuations are additive, and that the optimal revenue
	is $\sum_i 1/i = \Theta(\log m)$. Consider a uniform bundle price where $p^b(e) = 1/c$ for some $1 \leq c \leq m$. Then, the seller can sell edges that have valuation at least $1/c$. Observe that the number of such edges is at most $c$. Therefore, the revenue can be at most $\sum_{e:v_e \leq 1/c} 1/c = O(1)$.
\end{proof}

\lemitem*

\begin{proof}
	Let $\mC_i$ denote the class of customers that desire exactly $i $ items. We construct the hypergraph instance as follows: Each class of customers $\mC_i$ has 
	exactly size $\lceil n/i \rceil$ and each customer in $\mC_i$ is assigned a partition of $i$ items such that no two customers share any item. Thus, the total number of hyperedges is $m = \sum_{i} \vert \mC_i \vert = \Theta( n \log n)$. We fix the valuation $v_e = 1$ for all hyperedges. Selling each edge at price $p^{b}(e) = 1$, we extract the full revenue of $\Theta(n \log n)$.
	
	Next, we show that no item pricing solution can do better than $O(n)$. We will show this by induction on the customer class $\mC_i$. The base case is revenue obtained by selling edges to customers in $\mC_1$. Since there are at most $n$ edges, the maximum revenue is $O(n)$. Consider the customer class $\mC_k$. Let denote $0 < \alpha \leq 1$ be the fraction of edges sold to customers in $\mC_k$ and let $\mE_k$ denote such edges. Then, the maximum revenue that can be extracted is $\alpha \lceil n/k \rceil$. We distinguish three types of edges: $(i)$ edges that share at least one item with edges in $\mE_k$ $(ii)$ edges that do not share any item with edges in $\mE_k$ $(iii)$ edges that are strictly contained within $\mE_k$. By the induction hypothesis, the maximum revenue that can be extracted from all type $(ii)$ edges is at most $O(n)$. Each edge in $\mE_k$ can overlap with at most $2(k-1)$ edges that have size strictly lesser than $k$. Thus, the revenue extracted from type $(i)$ edges is at most $2(k-1) \alpha \lceil n/k \rceil = O(n)$. Consider an edge $e' \in \mE_k$ and let $w_1, \dots w_k$ be the weights of items in $e'$. Since the seller is able to sell $e'$, we have that $w_1 + \dots w_k \leq 1$. Observe that the maximum revenue of all edges of size strictly lesser than $k$ (say $k'$) using items in $e'$ is also at most $w_1 + \dots w_k \leq 1$ which gives revenue of $O(k)$, the revenue of all type $(iii)$ edges. Therefore, for each $i \in \{1, \dots, n\}$, the total revenue extracted from all  customer classes $C_1, \dots, C_i$ is at most $O(n)$. This completes the proof.
\end{proof}

\lemmax*

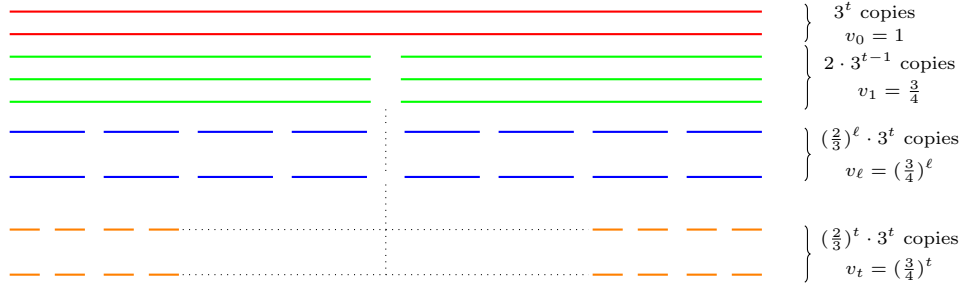
\begin{figure}[t] \centering
	\scalebox{1}{
		\begin{tikzpicture}
			\draw [red, thick] (-5,5) to (5,5);
			\draw [red, thick] (-5,4.7) to (5,4.7);
			
			\draw [decorate,decoration={brace,amplitude=2pt,raise=2pt},yshift=0pt]
			(5.5,5.1) -- (5.5,4.6) node [black,midway,xshift=1cm] {
				\shortstack{ \scriptsize $3^t$ copies \\ \scriptsize $v_0 = 1$}};
			
			\draw [green, thick] (-5,4.4) to (-0.2,4.4);
			\draw [green, thick] (0.2,4.4) to (5,4.4);
			
			\draw [green, thick] (-5,4.1) to (-0.2,4.1);
			\draw [green, thick] (0.2,4.1) to (5,4.1);			
			
			\draw [green, thick] (-5,3.8) to (-0.2,3.8);
			\draw [green, thick] (0.2,3.8) to (5,3.8);			
			
			\draw [decorate,decoration={brace,amplitude=2pt,raise=2pt},yshift=0pt]
			(5.5,4.55) -- (5.5,3.7) node [black,midway,xshift=1.2cm] {
			\shortstack{ \scriptsize $2 \cdot 3^{t-1}$ copies \\ \scriptsize $v_1 = \frac{3}{4}$}};			
		
			\draw[dotted]  (0, 3.7) -- (0, 3.45);
			
			\draw [blue, thick] (-5,3.4) to (-4,3.4);
			\draw [blue, thick] (-3.75,3.4) to (-2.75,3.4);			
			\draw [blue, thick] (-2.5,3.4) to (-1.5,3.4);			
			\draw [blue, thick] (-1.25,3.4) to (-0.25,3.4);									
			\draw [blue, thick] (5,3.4) to (4,3.4);
			\draw [blue, thick] (3.75,3.4) to (2.75,3.4);			
			\draw [blue, thick] (2.5,3.4) to (1.5,3.4);			
			\draw [blue, thick] (1.25,3.4) to (0.25,3.4);			
			
			\draw[dotted]  (0, 3.4) -- (0, 2.85);
			
			\draw [blue, thick] (-5,2.8) to (-4,2.8);
			\draw [blue, thick] (-3.75,2.8) to (-2.75,2.8);			
			\draw [blue, thick] (-2.5,2.8) to (-1.5,2.8);			
			\draw [blue, thick] (-1.25,2.8) to (-0.25,2.8);									
			\draw [blue, thick] (5,2.8) to (4,2.8);
			\draw [blue, thick] (3.75,2.8) to (2.75,2.8);			
			\draw [blue, thick] (2.5,2.8) to (1.5,2.8);			
			\draw [blue, thick] (1.25,2.8) to (0.25,2.8);
			
			\draw [decorate,decoration={brace,amplitude=2pt,raise=2pt},yshift=0pt]
			(5.5,3.45) -- (5.5,2.75) node [black,midway,xshift=1.2cm] {
				\shortstack{ \scriptsize $(\frac{2}{3})^\ell \cdot 3^{t}$ copies \\ \scriptsize $v_\ell = (\frac{3}{4})^{\ell}$}};			
			
			\draw[dotted]  (0, 2.7) -- (0, 2.1);
			
			\draw [orange, thick] (-5,2.1) to (-4.6,2.1);
			\draw [orange, thick] (-4.4,2.1) to (-4,2.1);
			\draw [orange, thick] (-3.75,2.1) to (-3.35,2.1);
			\draw [orange, thick] (-3.15,2.1) to (-2.75,2.1);			
			\draw[dotted] (-2.7, 2.1) to (2.7, 2.1)			;
			\draw [orange, thick] (5,2.1) to (4.6,2.1);
			\draw [orange, thick] (4.4,2.1) to (4,2.1);
			\draw [orange, thick] (3.75,2.1) to (3.35,2.1);
			\draw [orange, thick] (3.15,2.1) to (2.75,2.1);

			\draw[dotted]  (0, 2.1) -- (0, 1.5);
			
			\draw [orange, thick] (-5,1.5) to (-4.6,1.5);
			\draw [orange, thick] (-4.4,1.5) to (-4,1.5);
			\draw [orange, thick] (-3.75,1.5) to (-3.35,1.5);
			\draw [orange, thick] (-3.15,1.5) to (-2.75,1.5);			
			\draw[dotted] (-2.7, 1.5) to (2.7, 1.5)			;
			\draw [orange, thick] (5,1.5) to (4.6,1.5);
			\draw [orange, thick] (4.4,1.5) to (4,1.5);
			\draw [orange, thick] (3.75,1.5) to (3.35,1.5);
			\draw [orange, thick] (3.15,1.5) to (2.75,1.5);			
			
			\draw [decorate,decoration={brace,amplitude=2pt,raise=2pt},yshift=0pt]
			(5.5,2.15) -- (5.5,1.4) node [black,midway,xshift=1.2cm] {
				\shortstack{ \scriptsize $(\frac{2}{3})^{t} \cdot 3^{t}$ copies \\ \scriptsize $v_{t} = (\frac{3}{4})^{t}$}};

		\end{tikzpicture}
	}
	\caption{Laminar family construction for the lower bound of Lemma~\ref{lem:lb3}.}
	\label{fig:laminar}
\end{figure}

\begin{proof}

We construct a laminar family of sets arranged in binary tree fashion as follows: The root node is the set of all $n = 2^{t}$ items. At depth $\ell$, there are $2^\ell$ sets, each of size $\left( \frac{n}{2^{\ell}} \right)$ formed by partitioning each set at depth $\ell-1$ of size $\left( \frac{n}{2^{\ell - 1}} \right)$ into two. Further, each set $S$ at depth $\ell$ has valuation $v(S) = \left( \frac{3}{4} \right)^{\ell}$ and $c_\ell = \left( \frac{2}{3} \right)^{\ell} 3^t$ copies of itself. Figure~\ref{fig:laminar} shows the construction. Let $\mL$ denote the family of laminar sets in the construction. Although $v$ is only defined on $\mL$, it can be extended to any set $A\subseteq V$ by defining $v(A) = \min_{X \subseteq \mL} \sum_{S \in X} v(S)$ where $X$ is a set cover of $A$. Note that $v(A)$ is monotone and subadditive. We first prove that $v(A)$ is actually submodular.

Let us introduce some more notations. Let $\bA$ denote the minimum weighted set covering of a set $A$, i.e. $\bA = \argmin_{X \subseteq \mL} \sum_{S \in X} v(S)$  where $X$ is a set cover of $A$. We overload the valuation function $v$ to mean $v(\bA) = \sum_{S \in \bA} v(S)$ and define $A^{*} = \bigcup_{X \in \bA} X$ to be the set of items covered by some set in $X$. For any two sets $A, B \in 2^{[n]}$, it holds that $v(A) + v(B) = v(\bA) + v(\bB)$ by definition of valuation function. Observe that $A \cap B \subseteq A^{*} \cap B^{*}$ and by monotonicity, we have $v(A \cap B) \leq v(A^{*} \cap B^{*})$. Similarly, $v(A \cup B) \leq v(A^{*} \cup B^{*})$. Thus, to show submodularity, it suffices to verify that 
\begin{equation}\label{eqnsubmodular}
v(\bA) + v(\bB) \geq v(A^{*} \cup B^{*}) + v(A^{*} \cap B^{*}).
\end{equation}
Assume that the largest set in $\bA\cup \bB$ has size $2^u$. We verify (\ref{eqnsubmodular}) by applying induction to $u$.

\smallskip
\introparagraph{Base Case} If $u=0$, i.e. $\bA$ and $\bB$ only contain sets with size 1, let $L^{0}_1, \dots, L^{0}_{n} \in \mL$ denote all singleton sets. Define for each $i$ $A_i = A^* \cap L^{0}_i, B_i = B^* \cap L^{0}_i$. Then
\begin{eqnarray*}
	& &v(\bA) + v(\bB) \\
	&=& v(A_1) + \dots + v(A_n) +  v(B_1) + \dots + v(B_n) \\
	& = & v(A_1) + v(B_1) + \dots + v(A_n) + v(B_n) \\
	& = & v(A_1 \cap B_1) + \dots + v(A_n \cap B_n) + v(A_1 \cup B_1) + \dots + v(A_n \cup B_n) \\
	& = & v((A^* \cap B^*) \cap L^{0}_1) + \dots + v( (A^* \cap B^*) \cap L^{0}_n) +  v( (A^* \cup B^*) \cap L^{0}_1) + \dots + v( (A^* \cup B^*) \cap L^{0}_n) \\
	& \geq & v(A^{*} \cup B^{*}) + v(A^{*} \cap B^{*}).
\end{eqnarray*}
Here the first equality holds since $v(\bA)$ and $v(\bB)$ are defined only by sets with size 1.
The third equality holds since each $\bA_i$ and $\bB_i$ is either an empty set or singleton. The fourth equality holds by definition of $A_i$ and $B_i$. The fifth inequality holds by subadditivity of $v$.

\smallskip
\introparagraph{Inductive Case} Assume that $v(\bA) + v(\bB) \geq v(A^{*} \cup B^{*}) + v(A^{*} \cap B^{*})$ holds if all sets in $\bA\cup \bB$ have size at most $2^{u-1}$. Consider all sets of size $2^u$: $L^{u}_1, \dots, L^{u}_{n/2^u} \in \mL$ and define for every $i$ $A_i = A^* \cap L^{u}_i, B_i = B^* \cap L^{u}_i$. Using the same argument as in the base case, we have 

\begin{eqnarray*}
& &v(\bA) + v(\bB)\\
& = &v(A_1) + \dots + v(A_{n/2^{u}}) +  v(B_1) + \dots + v(B_{n/2^{u}})\\
& =  & v(A_1) + v(B_1) + \dots + v(A_{n/2^{u}}) + v(B_{n/2^{u}}) \\
& \geq & v(A_1 \cap B_1) + \dots + v(A_{n/2^{u}} \cap B_{n/2^{u}}) + v(A_1 \cup B_1) + \dots + v(A_{n/2^{u}} \cup B_{n/2^{u}}) \\
& = & v( (A^* \cap B^*) \cap L^{u}_1) + \dots + v( (A^* \cap B^*) \cap L^{u}_{n/2^{u}}) +  v( (A^* \cup B^*) \cap L^{u}_1) + \dots + v( (A^* \cup B^*) \cap L^{u}_{n/2^{u}}) \\
&\geq &v(A^{*} \cup B^{*}) + v(A^{*} \cap B^{*}).
\end{eqnarray*}

Here the third inequality is where we invoke the inductive hypothesis: if $v(A_i)=v(L_i^u)$ or $v(B_i)=v(L_i^u)$, then $v(A_i)+v(B_i)=v(A_i\cup B_i)+v(A_i\cap B_i)$. Otherwise, $v(A_i)$ and $v(B_i)$ are both defined by sets with size at most $2^{u-1}$, by inductive hypothesis $v(A_i)+v(B_i)\geq v(A_i\cup B_i)+v(A_i\cap B_i)$. Thus (\ref{eqnsubmodular}) always holds, which means $v$ is indeed a submodular function.

If we price every bundle at its value, the revenue extracted is maximized. For each level $\ell$, there are $2^{\ell}$ sets and $(2/3)^{\ell} \cdot 3^{t}$ copies of each set. Thus, the total revenue from all $t+1$ levels is
\begin{align*}
	\textsf{OPT} = \sum_{\ell = 0}^{t} v_\ell c_\ell 2^\ell = (t+1)\left( \frac{3}{4} \right)^{\ell}  \cdot \left( \frac{4}{3} \right)^{\ell} \cdot  3^{t} = (t+1) \cdot 3^{t}.
\end{align*}




Next, we will show that no uniform bundle pricing or item pricing can extract revenue more than $O(3^{t})$.

For optimal bundle pricing, the first observation is that we need to consider only bundle prices of the form $\left( \frac{3}{4} \right)^{k}$, i.e. the value of some set in the laminar system. Let the bundle price chosen be $\left(\frac{3}{4} \right)^{k}$, then we can sell all edges at depth $\leq k$. The revenue collected by such bundle pricing is 
\begin{equation*}
\textsf{OPT}_B = \sum_{i \leq k} \left( \frac{3}{4} \right)^{k} \cdot \left( \frac{2}{3} \right)^{i} \cdot 3^{t} \cdot 2^{i} 
 = \left( \frac{3}{4} \right)^{k} \cdot 3^{t} \sum_{i \leq k} \left( \frac{4}{3} \right)^{i} \displaybreak[0]\\
 \leq \left( \frac{3}{4} \right)^{k} \cdot 3^{t+1} \cdot \left( \frac{4}{3} \right)^{k} = O(3^{t}).
\end{equation*}

For optimal item pricing, let $k$ be the smallest depth for any set sold by an optimal item pricing solution. By symmetry we can assume that all sets with depth $k$ are sold. Then the sum of prices of all items is upper bounded by $\sum_{i}p_i\leq (\frac{3}{4})^{k}\cdot 2^k=(\frac{3}{2})^k$, since there are $2^k$ distinct sets with depth $k$. Thus the revenue of such item pricing is contributed by sets with depth at least $k$, and can be upper bounded by
\begin{equation*}
\textsf{OPT}_{IP}\leq\sum_{\ell\geq k}\left(\frac{3}{2}\right)^k\cdot c_\ell=\left( \frac{3}{2} \right)^{k} \cdot   3^{t} \sum_{\ell \geq k} \left( \frac{2}{3} \right)^{\ell}
\leq \left( \frac{3}{2} \right)^{k} \cdot  3^{t+1} \cdot  \left( \frac{2}{3} \right)^{k} 
=O(3^{t})
\end{equation*}
since each set in level $\ell$ has $c_\ell$ copies.

In conclusion the revenue gap between $\textsf{OPT}$ and $\max(\textsf{OPT}_B,\textsf{OPT}_{IP})$ is $\Omega(t)$. Note that $m = \sum_{k = 0}^{t} 3^{t} \cdot  2^{k} \cdot  \left( \frac{2}{3} \right)^{k} \leq 3^{t} \cdot  \left( \frac{4}{3} \right)^{t+1} = O(4^{t})$ and thus, the revenue gap is $\Omega(\log m)$.

\end{proof}

\begin{table} \centering
	\scalebox{0.90}{
		\begin{tabular}{l|L{12cm}}
			\toprule
			& \textbf{Query}  \\
			\midrule
			
			$Q_1$ & \texttt{
				select count(Name) from Country where Continent = `Asia'
			} \\ \hdashline 
			
			$Q_2$ & \texttt{
				select count(distinct Continent) from Country
			} \\ \hdashline
			
			$Q_3$ & \texttt{
				select avg(Population) from Country 
			} \\ \hdashline
			
			$Q_4$ & \texttt{
				select max(Population) from Country
			} \\ \hdashline
			
			$Q_5$ & \texttt{
				select min(LifeExpectancy) from Country
			} \\ \hdashline
			
			$Q_6$ & \texttt{
				select count(Name) from Country where Name like `A
			} \\ \hdashline
			
			$Q_7$ & \texttt{
				select Region, max(SurfaceArea) from Country group by Region 
			} \\ \hdashline
			
			$Q_8$ & \texttt{
				select Continent, max(Population) from Country group by Continent
			} \\ \hdashline
			
			$Q_9$ & \texttt{
				select Continent, count(Code) from Country group by Continent
			}\\

			$Q_{10}$ & \texttt{
				select * from Country
			} \\ \hdashline
			
			$Q_{11}	$ & \texttt{
				select Name from Country where Name like 'A\%'
			} \\ \hdashline
			
			$Q_{12}$ & \texttt{
				select * from Country where Continent='Europe' and Population > 5000000 
			} \\ \hdashline
			
			$Q_{13}$ & \texttt{
				select * from Country where Region='Caribbean' 
			} \\ \hdashline
			
			$Q_{14}$ & \texttt{
				select Name from Country where Region='Caribbean'
			}\\
			
			$Q_{15}$ & \texttt{
				select Name from Country where Population between 10000000 and 20000000
			} \\ \hdashline
			
			$Q_{16}$ & \texttt{
				select * from Country where Continent='Europe' limit 2
			} \\ \hdashline
			
			$Q_{17}$ & \texttt{
				select Population from Country where Code = USA'
			} \\ \hdashline
			
			$Q_{18}$ & \texttt{
				select GovernmentForm from Country
			} \\ \hdashline
			
			$Q_{19}$ & \texttt{
				select distinct GovernmentForm from Country
			} \\ \hdashline
			
			$Q_{20}$ & \texttt{
				select * from City where Population >= 1000000 and CountryCode = 'USA' 
			} \\ \hdashline
			
			$Q_{21}$ & \texttt{
				select distinct Language from CountryLanguage where CountryCode='USA'
			} \\ \hdashline
			
			$Q_{22}$ & \texttt{
				select * from CountryLanguage where IsOfficial = 'T'
				'} \\ \hdashline
			
			$Q_{23}$ & \texttt{
				select Language, count(CountryCode) from CountryLanguage group by Language
			} \\ \hdashline
			
			$Q_{24}$ & \texttt{
				select count(Language) from CountryLanguage where CountryCode = 'USA'
			} \\ \hdashline
			
			$Q_{25}$ & \texttt{
				select CountryCode, sum(Population) from City group by CountryCode 
			} \\ \hdashline
			
			$Q_{26}$ & \texttt{
				select CountryCode, count(ID) from City group by CountryCode 
			} \\ \hdashline
			
			$Q_{27}$ & \texttt{
				select * from City where CountryCode = 'GRC'
			} \\ \hdashline
			
			$Q_{28}$ & \texttt{
				select distinct 1 from City where CountryCode = 'USA' and Population > 10000000
			} \\ \hdashline
			
			$Q_{29}$ & \texttt{
				select Name from Country , CountryLanguage where Code = CountryCode and Language = 'Greek'
			} \\ \hdashline
			
			$Q_{30}$ & \texttt{
				select C.Name from Country C, CountryLanguage L where C.Code = L.CountryCode and L.Language = 'English' and L.Percentage >= 50
			} \\ \hdashline
			
			$Q_{31}$ & \texttt{
				select T.district from Country C, City T where C.code = 'USA' and C.capital = T.id
			} \\ \hdashline
			
			$Q_{32}$ & \texttt{
				select * from Country C, CountryLanguage L where C.Code = L.CountryCode and L.Language = 'Spanish'
			} \\ \hdashline
			
			$Q_{33}$ & \texttt{
				select Name, Language from Country , CountryLanguage where Code = CountryCode 
			} \\ \hdashline
			
			$Q_{34}$ & \texttt{
				select * from Country , CountryLanguage where Code = CountryCode
			} \\ \hdashline
		\end{tabular}
	}
	\vspace{2mm}
	\caption{Queries used for \texttt{world} database}
	\label{table:worldqueries}
\end{table}

\section{Skewed Workload}
\label{sec:skewed:workload}

Table~\ref{table:worldqueries} lists the queries used for the skewed workload on $\texttt{\bfseries world}$ dataset. To increase the number of queries to $986$, we add a new query for every country (in the domain of the attribute) by changing the predicate in $Q_{17}, Q_{27}, Q_{31}$, for every continent in $Q_1, Q_{12}$ and for every language in $Q_{29}, Q_{30}$.

\section{Generating  \tpch\ and \ssb\ workload}
\label{sec:tpch:ssb}

\introparagraph{\tpch} $Q_4, Q_6, Q_{12}, Q_1$ are parameterized by \texttt{year} whose active domain is $[1993, \dots, 1998]$. This gives $20$ queries. Query $Q_2$ generates $5$ queries, one for each region. $Q_{16}$ gives $150$ queries parameterized for $150$ \texttt{p\_type} values. $Q_{17}$ generate $40$ queries by parameterizing \texttt{p\_container}. $Q_2$ is parameterized by \texttt{p\_type} by the active domain $[$\texttt{BRASS, TIN, COPPER, STEEL, NICKEL}$]$.

\smallskip
\noindent \introparagraph{\ssb} Similar to \tpch\ , the parameters are \texttt{year} ($7$), \texttt{region} ($5$), \texttt{nation} ($25$), \texttt{city} ($250$). $Q_1, Q_2, Q_3$ generate one query for each \texttt{year},  $Q_4, Q_5, Q_6, Q_7, Q_{11, Q_{12}}$ generate one query for each \texttt{region},  $Q_9, Q_{10}$ generate one query for each \texttt{city} and $Q_{42}$ creates one query for each $(\mathtt{region}, \mathtt{nation})$ pair.